\documentclass[%
 reprint,
 superscriptaddress,
 amsmath,amssymb,
 mathtools,
 aps, 
 pra,
]{revtex4-2}

\usepackage{graphicx}
\usepackage[colorlinks=true,allcolors=blue]{hyperref}
\usepackage{diffcoeff}

\newcommand{\abs}[1]{\vert #1\vert}  
\newcommand{\mean}[1]{\langle #1 \rangle}

\newcommand{\kB}{k_\text{B}}

\newcommand{\m}{\text{m}}
\newcommand{\Om}{\Omega_\m}
\newcommand{\omL}{\omega_\text{las}}
\newcommand{\aL}{\alpha_\text{las}}
\newcommand{\omd}{\omega_d}
\newcommand{\oma}{\omega_a}
\newcommand{\gd}{\gamma_d}

\newcommand{\e}{\text{e}}
\renewcommand{\L}{1}
\newcommand{\R}{2}

\newcommand{\Nm}{\bar{n}_\m}

\newcommand{\Pin}{{P}_\text{las}}

\newcommand{\ain}{\hat{a}_{\text{in},\mu}}
\newcommand{\ainL}{\hat{a}_{\text{in,}\L}}
\newcommand{\ainR}{\hat{a}_{\text{in,}\R}}

\newcommand{\aoutR}{\hat{a}_{\text{out,}\R}}

\newcommand{\da}{\delta\hat{a}}
\newcommand{\dq}{\delta\hat{q}}
\newcommand{\deltp}{\delta\hat{p}}  
\newcommand{\deltd}{\delta\hat{d}}  

\newcommand*{\gam}{\gamma_\m}
\newcommand*{\kL}{\kappa_\L}
\newcommand*{\kR}{\kappa_\R}
\newcommand*{\G}{\mathcal{G}}
\newcommand*{\Da}{\Delta_a}
\newcommand*{\Dd}{\Delta_d}

\newcommand*{\gw}[1][c]{g_{\omega,#1}}
\newcommand*{\gk}[1][c]{g_{\kappa,#1}}
\newcommand*{\gwa}{\gw[a]}
\newcommand*{\gwd}{\gw[\smash{d}]}
\newcommand*{\gka}{\gk[a]}
\newcommand*{\gkd}{\gk[\smash{d}]}
\newcommand*{\gadm}{g_{a,\smash{d}}^{0, -}}
\newcommand*{\gadp}{g_{a,\smash{d}}^{0, +}}

\newcommand*{\tg}[1][c]{\tilde{g}_{#1}}
\newcommand*{\tga}{\tg[a]}
\newcommand*{\tgd}{\tg[d]}
\newcommand*{\tgm}[1][c]{\tilde{g}_{\m,#1}}
\newcommand*{\tgma}{\tgm[a]}
\newcommand*{\tgmd}{\tgm[d]}

\newcommand{\eref}[1]{Eq.~\ref{#1}}
\newcommand{\fref}[1]{Fig.~\ref{#1}}

\newcommand{\sref}[1]{Sec.~\ref{#1}}
\newcommand{\PhC}{\text{PhC}}

\newcommand{\SM}{Supplementary Material}

\newcommand{\beginsupplement}{%
        \setcounter{section}{0}
        \setcounter{table}{0}
        \renewcommand{\thetable}{S\arabic{table}}%
        \setcounter{figure}{0}
        \renewcommand{\thefigure}{S\arabic{figure}}%
        \renewcommand{\theHfigure}{S\arabic{figure}}
        \setcounter{equation}{0}
        \renewcommand{\theequation}{S\arabic{equation}}%
                \renewcommand{\thesection}{S\arabic{section}}%
     }

\begin{document}

\title{Integrated microcavity optomechanics with a suspended photonic crystal mirror above a distributed Bragg reflector}

\author{Sushanth Kini Manjeshwar}
\thanks{Equal author contribution}
\affiliation{%
Department of Microtechnology and Nanoscience (MC2), Chalmers University of Technology, SE-412 96 G{\"o}teborg, Sweden\\
 }
 \author{Anastasiia Ciers}
 \thanks{Equal author contribution}
\affiliation{%
Department of Microtechnology and Nanoscience (MC2), Chalmers University of Technology, SE-412 96 G{\"o}teborg, Sweden\\
 }
\author{Juliette Monsel}
\thanks{Equal author contribution}
\affiliation{%
Department of Microtechnology and Nanoscience (MC2), Chalmers University of Technology, SE-412 96 G{\"o}teborg, Sweden\\
 }
 \author{Hannes Pfeifer}
\affiliation{%
Department of Microtechnology and Nanoscience (MC2), Chalmers University of Technology, SE-412 96 G{\"o}teborg, Sweden\\
 }
\author{Cindy Peralle}
\affiliation{%
Department of Physics, Chalmers University of Technology, SE-412 96 G{\"o}teborg, Sweden\\
 }
\author{Shu Min Wang}
\affiliation{%
Department of Microtechnology and Nanoscience (MC2), Chalmers University of Technology, SE-412 96 G{\"o}teborg, Sweden\\
 }
\author{Philippe Tassin}
\affiliation{%
Department of Physics, Chalmers University of Technology, SE-412 96 G{\"o}teborg, Sweden\\
 }
 \author{Witlef Wieczorek}
\email{witlef.wieczorek@chalmers.se}
\affiliation{%
Department of Microtechnology and Nanoscience (MC2), Chalmers University of Technology, SE-412 96 G{\"o}teborg, Sweden\\
 }%

\date{\today}

\begin{abstract}

Increasing the interaction between light and mechanical resonators is an ongoing endeavor in the field of cavity optomechanics. Optical microcavities allow for boosting the interaction strength through their strong spatial confinement of the optical field. In this work, we follow this approach by realizing a sub-wavelength-long, free-space optomechanical microcavity on-chip fabricated from an (Al,Ga)As heterostructure. A suspended GaAs photonic crystal mirror is acting as a highly reflective mechanical resonator, which together with a distributed Bragg reflector forms an optomechanical microcavity. We demonstrate precise control over the microcavity resonance by change of the photonic crystal parameters. The interplay between the microcavity mode and a guided resonance of the photonic crystal modifies the cavity response and results in a stronger dynamical backaction on the mechanical resonator compared to conventional optomechanical dynamics. 

\end{abstract}

\maketitle

\section{Introduction}

Optical microcavities \cite{vahala_optical_2003} confine light to small mode volumes leading to stronger light-matter interactions. As a consequence, microcavities have been used in various fields, including cavity QED \cite{najer_gated_2019,Albrecht2013,reithmaier_strong_2004}, nonlinear optics \cite{Lin:17,Kippenberg2004}, or vertical-cavity surface-emitting lasers \cite{michalzik_vcsels_2013, Lina2013}. Microcavities also find applications in the field of cavity optomechanics \cite{Aspelmeyer2014Dec}, which explores the interaction between a cavity mode and a mechanical resonator. Optomechanical microcavities have enabled demonstrations such as ground-state cooling of mechanical motion \cite{chan_laser_2011,teufelSidebandCoolingMicromechanical2011}, coherent optomechanical coupling 
\cite{groblacherObservationStrongCoupling2009,verhagen_quantum-coherent_2012}, and thermally driven nonlinear optomechanics \cite{leijssen_nonlinear_2017}. 

Optomechanical microcavities can be categorized into two different topologies. Firstly, light can be confined fully within the material, i.e., in the plane of the device layer, as is the case with optomechanical crystals  \cite{Jiang2009,eichenfield_optomechanical_2009}, whispering gallery mode resonators \cite{Ding2010,Hofer2010}, or photonic crystal defect cavities \cite{tsvirkunIntegratedIIIVPhotonic2015}. Alternatively, light can be confined in free space, i.e., out-of-plane, by use of highly reflecting mirrors such as in Fabry-Pérot-type microcavities \cite{wachter_silicon_2019,Hornig:20,pfeifer_FPMicrocavity_2022}. While the former implementations allow for on-chip integration, inherent mechanical stability, and microfabrication flexibility, they may suffer from optical absorption \cite{Meenehan2014}. The latter implementations circumvent this caveat but lack microfabrication flexibility and intrinsic mechanical stability. 

In our work, we realize a high-quality optomechanical microcavity by combining in-plane and out-of-plane light propagation. We confine light to a sub-$\mu$m microcavity mode formed between a distributed Bragg reflector (DBR) and a suspended photonic crystal (PhC) mirror. Our PhC-DBR microcavity combines the advantages of inherent mechanical stability realized through a single-lithography microfabrication process from an (Al,Ga)As heterostructure \cite{kini2020suspended} with on-chip scalability and flexibility over design parameters. The reflection of the PhC mirror is determined by the Fano interference between out-of-plane and in-plane light modes and can be engineered to obtain a reflectivity approaching unity \cite{fan2002}. At the same time, the suspended PhC mirror acts as the mechanical resonator \cite{makles_2d_2015,kini2020suspended,Moura:18,Catavu2012} of the optomechanical cavity. Suspended PhC mirrors have been used to realize cavity optomechanical systems, such as for an optomechanical microcavity exhibiting both photothermal and optomechanical effects \cite{rodriguezBondingAntibondingTunable2011,Woolf:13}, as membrane-in-the-middle cavity optomechanical systems \cite{chen_high-finesse_2017,Claus2018}, as photonic crystal cavities \cite{zhouCavityOptomechanicalBistability2022,enzianPhononicallyShieldedPhotoniccrystal2023} or defect cavities \cite{tsvirkunIntegratedIIIVPhotonic2015}. 
Furthermore, suspended PhC mirrors exhibit features for cavity optomechanical systems, which are not accessible by conventional mirrors, such as focusing capability \cite{Guo2017}, linewidth narrowing \cite{Naesby2018}, or realization of photonic bound states in the continuum \cite{fitzgerald2021cavity}. 

In our PhC-based optomechanical microcavity, we observe a stronger dynamical backaction compared to canonical cavity optomechanics. This stronger interaction originates from the interplay between the microcavity mode and a guided resonance of the photonic crystal \cite{fan2002}, which was analyzed in Refs.~\cite{Naesby2018,Cernotik2019Jun}. We extend that analysis by incorporating dispersive coupling of the guided resonance of the suspended PhC to its mechanical motion. Moreover, we also account for dissipative coupling of the optical modes to the motion of the suspended PhC \cite{xuerebDissipativeOptomechanicsMichelsonSagnac2011,tsvirkunIntegratedIIIVPhotonic2015,baraillonLinearAnalyticalApproach2020}.

\begin{figure*}[t!bhp]
    \centering\includegraphics{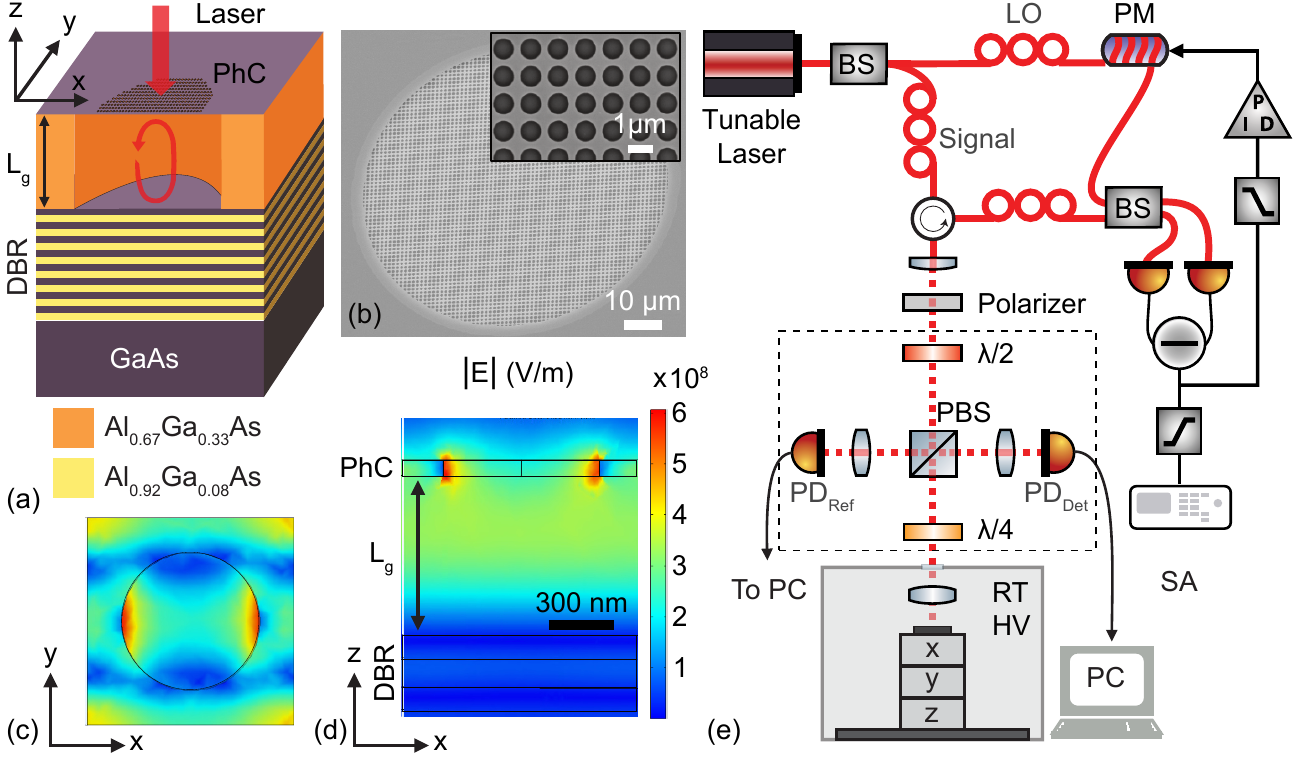}
    \caption{(a) Schematic of a microcavity formed by a suspended photonic crystal (PhC) mirror and a distributed Bragg reflector (DBR) mirror monolithically fabricated from an (Al,Ga)As heterostructure. (b) SEM image of a suspended PhC membrane with a diameter of 70\,$\mu$m. (c,d) FEM simulation of the electric field $|E|$ of the microcavity mode at a resonant wavelength of 1497\,nm. (e) Experimental setup used for characterizing the optomechanical properties of the microcavity. Red solid lines represent fiber beam paths and red dashed lines free-space beam paths. Dashed rectangle: optional reflection measurement setup, BS: beam splitter, PBS: polarizing beam splitter, PM: fiber-based phase modulator, SA: spectrum analyzer, PD$_\mathrm{{Ref(Det)}}$: photodiode in the reference (detection) arm, RT: room temperature, HV: high vacuum of $5\cdot 10^{-5}\,\mathrm{mbar}$.}
    \label{fig:sem_1}
\end{figure*}

In the following, we present the fabrication of the optomechanical microcavity from an (Al,Ga)As heterostructure. We summarize the theory required to describe the optical and optomechanical properties of our microcavity. We then characterize the optical reflectivity of the microcavity and demonstrate the tunability of the cavity resonance wavelength by variation of the PhC hole radius. Finally, we characterize the mechanical properties of the suspended PhC reflector and their tunability through the use of cavity optomechanical effects, which deviate considerably from canonical optomechanics \cite{Aspelmeyer2014Dec}. 


\section{Experimental Methods}
\subsection{Device Fabrication}

The optomechanical microcavity is fabricated from an (Al,Ga)As heterostructure, which is grown on a GaAs substrate using molecular beam epitaxy. This monolithic assembly avoids post-alignment of optical mirrors forming a microcavity and also active cavity length stabilization, which is required, for instance, in fiber-based microcavities \cite{Hunger2010,Peng2019}. The as-grown heterostructure [\fref{fig:sem_1}(a)] consists of a DBR mirror with 30 alternating layers of 106\,nm-thick GaAs and 130\,nm-thick Al$_{0.92}$Ga$_{0.08}$As with a targeted reflectivity of 99.999\% at 1480\,nm. The DBR is followed by a 750\,nm-thick Al$_{0.67}$Ga$_{0.33}$As sacrificial layer, which determines the geometric length $L_g$ of the microcavity. On top of this sacrificial layer, a 95\,nm-thick GaAs layer is grown as the device layer (thicknesses of the as-grown structure are inferred from transmission electron microscopy, see \SM). The device layer is patterned with a PhC to increase its out-of-plane reflectivity in the telecom wavelength regime \cite{fan2002,makles_2d_2015,kini2020suspended}. Standard microfabrication techniques were used to pattern the PhC into the GaAs device layer and to release it \cite{kini2020suspended}. A scanning electron microscope (SEM) image of a fabricated device is shown in \fref{fig:sem_1}(b). The suspended PhC membrane together with the DBR mirror form a microcavity with a length $\approx \lambda/2$. \fref{fig:sem_1}(c,d) present the electric field distribution of the microcavity eigenmode obtained from a finite element method (FEM) simulation that clearly shows the field concentration between the PhC and DBR mirrors.

\subsection{Experimental setup}

The characterization of the microcavity is performed with a tunable diode laser in the telecom wavelength regime (1480\,nm to 1620\,nm) using the setup shown in \fref{fig:sem_1}(e). We use a standard optical homodyne detection scheme \cite{barg2017measuring} at room temperature to characterize the (opto)mechanical properties of the suspended PhC membrane placed in a vacuum chamber ($5\cdot10^{-5}$\,mbar). The output of the laser is split into a signal and a local oscillator (LO) arm. The signal arm is collimated to a free-space beam path and focused onto the microcavity inside a vacuum chamber on a \textit{xyz}-translation stage. The mechanical displacement of the suspended PhC membrane is mapped onto the quadratures of the reflected light beam. The reflected beam is mixed with the LO beam in a tunable fiber beam splitter, whose output is sent to a balanced photodiode (BPD). The electronic signal from the BPD is passed through a high-pass filter and sent to a spectrum analyzer for evaluating the mechanical properties of the PhC membrane. The low-pass filtered BPD signal is used to lock the interferometer by applying feedback to a fiber-based phase modulator. The reference photodiode PD$_\mathrm{Ref}$ is used to maintain a constant power reaching the sample when tuning the laser wavelength.

For characterizing the optical reflectance of the microcavity, additional components can be introduced into the free-space beam path of the signal arm (marked by the dashed box in \fref{fig:sem_1}(e)). Then, a circularly polarized beam acquires a $\pi$-phase shift upon reflection on the microcavity and, thus, will be directed to PD$_\mathrm{Det}$ after passing through a quarter-wave plate and polarizing beam splitter. To account for the optical response of the utilized components, we perform an independent calibration measurement with a mirror of known reflectivity that is used to normalize the obtained optical reflectance spectra of the microcavity \cite{kini2020suspended}. 


\section{Theory}

We briefly present the theoretical description of the optomechanical microcavity via coupled mode theory, which allows us to describe its optical and optomechanical properties. More details are found in the \SM{} and in Ref.~\cite{theory}.

\begin{figure}[t!bhp]
    \centering
    \includegraphics{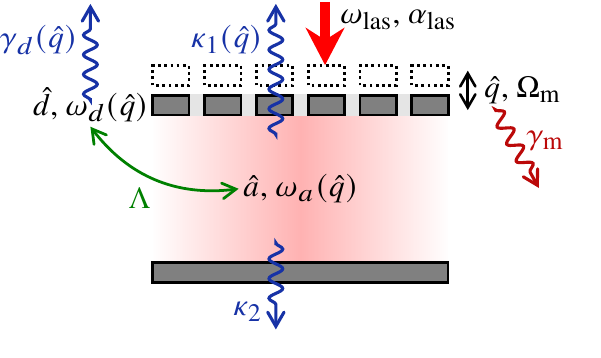}
    \caption{Schematic of the optomechanical microcavity consisting of a suspended PhC mirror above a DBR. Symbols are explained in the main text.}
    \label{fig:fig3}
\end{figure}

The system, depicted in \fref{fig:fig3}, consists of a suspended PhC mirror and a DBR mirror that form an optical microcavity. The DBR can be treated as a highly reflective mirror in the telecom range. Conversely, the optical response of the PhC mirror is highly frequency dependent, which we capture by accounting for a PhC guided resonance. Therefore, we model the microcavity with two coupled optical modes \cite{Cernotik2019Jun}: a Fabry-Pérot cavity mode of frequency $\oma$ with photon annihilation operator $\hat{a}$ and an internal PhC optical mode of frequency $\omd$ with photon annihilation operator $\hat{d}$. The Hamiltonian for the optical part of the setup reads

\begin{equation}
     \hat{H}_\text{opt} = \hbar\oma \hat{a}^\dagger \hat{a}+\hbar\omd \hat{d}^\dagger \hat{d} + \hbar\Lambda(\hat{a}^\dagger\hat{d} + \hat{a}\hat{d}^\dagger),
\end{equation}
where $\Lambda$ is the coupling strength between the modes and where we made the rotating wave approximation and neglected the two-mode squeezing terms \cite{Cernotik2019Jun}. In addition, the cavity mode couples to the environments above (denoted with subscript 1) and below (denoted with subscript 2) the cavity, giving rise to the respective loss rates $\kL$ and $\kR$. The mirror mode also couples to environment 1 with loss rate $\gd$ \cite{Cernotik2019Jun}.

The suspended PhC membrane acts as a mechanical resonator of frequency $\Om$ and position and momentum quadratures $\hat{q}$ and $\hat{p}$. The  displacement of the suspended PhC membrane impacts the optical microcavity in various ways, resulting in different kinds of optomechanical couplings. First, the modulation of the Fabry-Pérot cavity resonance frequency by the out-of-plane displacement of the PhC mirror creates the standard dispersive coupling, $\gwa^0$. Second, we expect a dissipative coupling $\gka^0$ of the Fabry-Pérot mode originating from the strong wavelength dependence of the PhC reflectivity. In other words, the dispersive coupling results in a new microcavity resonance frequency at which the reflectivity of the PhC is different compared to the initial one, thus, leading to a modulation of the loss rate $\kL$. Furthermore, we also take into account optomechanical couplings from the interaction of the mechanical mode of the PhC with its guided optical mode $\hat{d}$. The out-of-plane displacement of the PhC results in an induced in-plane mechanical strain and displacement field impacting the PhC properties. As a result, the PhC mode $\hat{d}$ will have an altered frequency $\omega_d$ and linewidth $\gamma_d$. We model this by adding dispersive and dissipative optomechanical couplings between the PhC mode and the mechanics, respectively denoted  $\gwd^0$ and $\gkd^0$. We note that we do not model quadratic coupling of the mechanical quadratures to the light field as we do not observe a mechanical signal at twice the frequency in the measured mechanical noise power spectra.

As a result, the total Hamiltonian of the system is given by
\begin{equation}
    \hat{H} = \hat{H}_\text{opt} + \frac{\hbar\Om}{2}(\hat{q}^2 + \hat{p}^2) -\hbar\gwa^0\sqrt{2}\hat{a}^\dagger \hat{a}\hat{q}-\hbar\gwd^0\sqrt{2}\hat{d}^\dagger \hat{d}\hat{q}.
\end{equation}
Furthermore, the cavity is driven by a laser of frequency $\omL$ and amplitude $\aL$, and the mechanical resonator is coupled to a phonon bath, giving rise to the damping rate $\gam$, so the dynamics of the whole system can be described, in the frame rotating at the laser frequency, by the following set of Langevin equations\cite{Cernotik2019Jun, Elste2009May, theory},
\begin{widetext}
\begin{align}\label{eq:Langevin eqs}
    \dot{\hat{a}}=\,& -(i\Delta_a + \kappa) \hat{a} + \sqrt{2}(i \gwa^0 - \gka^0)\hat{q}\hat{a}-\G \hat{d}- \sqrt{2}\gadp \hat{q}\hat{d}\\\nonumber
    &+ \left(\sqrt{2\kL} + \frac{\gka^0}{\sqrt{\kL}} \hat{q}\right)(\aL^{} + \ainL^{}) + \sqrt{2\kR}\ainR^{},\\\nonumber
     \dot{\hat{d}}=\,& -(i\Delta_d + \gd) \hat{d} +\sqrt{2}(i \gwd^0 - \gkd^0)\hat{q}\hat{d} -\G \hat{a} - \sqrt{2}\gadp\hat{q}\hat{a} + \left(\sqrt{2\gd} + \frac{\gkd^0}{\sqrt{\gd}} \hat{q}\right)(\aL^{} + \ainL^{}),\\\nonumber
     \dot{\hat{q}}=\,& \Om \hat{p},\\\nonumber
    \dot{\hat{p}}=\,& -\Om\hat{q} -\gam \hat{p}  + i\sqrt{2}\gadm(\hat{a}^\dagger\hat{d}- \hat{d}^\dagger\hat{a}) + \sqrt{2}\gwa^0  \hat{a}^\dagger \hat{a}  + \sqrt{2}\gwd^0  \hat{d}^\dagger \hat{d}  + \sqrt{\gam}\hat{\xi}\\\nonumber
    &- i\frac{\gka^0}{\sqrt{\kL}}(\hat{a}^\dagger (\aL^{} + \ainL^{}) - (\aL^* + \ainL^\dagger) \hat{a}) - i\frac{\gkd^0}{\sqrt{\gd}}(\hat{d}^\dagger (\aL^{} + \ainL^{}) - (\aL^* + \ainL^\dagger) \hat{d}).
\end{align}
\end{widetext}
We have defined the detunings $\Da = \oma - \omL$ and $\Dd = \omd - \omL$, the generalized coupling $\G = i\Lambda + \sqrt{\kL\gd}$, the total cavity loss rate $\kappa = \kL + \kR$ and the effective optomechanical couplings 
\begin{equation}\label{g_ad}
    g_{a,d}^{0,\pm} = \frac{\sqrt{\kL\gd}}{2}\left(\frac{\gka^0}{\kL} \pm \frac{\gkd^0}{\gd} \right).
\end{equation} 
At room temperature, we can neglect the thermal fluctuations in the optical environments, therefore $\ain$ corresponds to the vacuum input noise from environment $\mu = 1,2$ and its only non-zero correlation function is $\mean{\ain^{}(t)\ain^\dagger(t')} = \delta(t - t')$. Conversely, the mechanical environment is at high temperature, and the correlation function of the input mechanical noise $\hat{\xi}$ can be approximated by $\mean{\hat{\xi}(t)\hat{\xi}(t')} = (2\Nm + 1)\delta(t-t')$, where $\Nm = (\exp(\hbar\Om/\kB T) - 1)^{-1}$ is the average  phonon number in the mechanical environment.

Our microcavity is in the regime $\gwa^0, \gwd^0, \gka^0, \gkd^0 \ll \kappa, \gd, \Lambda$ (see results section), so we can neglect the optomechanical effects when modeling the optical response of the cavity and, like in Ref.~\cite{Cernotik2019Jun}, we obtain the transmission coefficient of the cavity
    \begin{equation}
        t = \frac{\sqrt{2\kR}\left(\sqrt{2\gd}\G - \sqrt{2\kL} (\gd + i\Dd)\right)}{(\kappa + i\Da)(\gd + i\Dd) - \G^2 }.
        \label{eq:COM_transmission}
    \end{equation}
Furthermore, the cavity and PhC modes are in the strong coupling regime, $\Lambda > \gd, \kappa$. We can thus estimate the effective optical resonances by diagonalizing the Langevin equations for the average optical modes, in the absence of the mechanical resonator,
\begin{equation}
    \begin{pmatrix}
        \mean{\dot{\hat{a}}}\\ \mean{\dot{\hat{d}}}
    \end{pmatrix}
    = 
    -i\begin{pmatrix}
        \oma - i\kappa & -i\G\\ -i\G & \omd - i\gd
    \end{pmatrix}
    \begin{pmatrix}
        \mean{{\hat{a}}}\\ \mean{{\hat{d}}}
    \end{pmatrix}
    +
    \begin{pmatrix}
        \sqrt{2\kL}\aL\\ \sqrt{2\gd}\aL
    \end{pmatrix}.
\end{equation}
We obtain the complex eigenvalues
\begin{equation}\label{eq:eigenfreq}
    \tilde\omega_\pm =\frac{\oma + \omd}{2} - i\frac{\kappa + \gd}{2} \pm \sqrt{\left(\frac{\oma - \omd}{2}-i\frac{\kappa - \gd}{2}\right)^2-\G^2 },
\end{equation}
which correspond to the effective resonance frequencies $\omega_\pm = \Re(\tilde{\omega}_\pm)$ and loss rates $\kappa_\pm = -\Im(\tilde{\omega}_\pm)$. Since the optomechanical effects can be treated as a perturbation, we can simply introduce the mechanical position dependency of $\oma$, $\omd$, $\kL$ and $\gd$ in Eq.~\eqref{eq:eigenfreq} and estimate the effective dispersive and dissipative couplings
\begin{equation}\label{eq:OM_eff_couplings}
    g_{\omega, \pm}^0 = -\frac{1}{\sqrt{2}}\diffp{\omega_\pm}{q}[q=0],\quad g_{\kappa, \pm}^0 = \frac{1}{\sqrt{2}}\diffp{\kappa_\pm}{q}[q=0].
\end{equation}

Finally, given our laser power, the set of equations \eqref{eq:Langevin eqs} can be linearized around a semiclassical steady-state, allowing us to solve them and derive the mechanical frequency shift due to the optomechanical interactions, see \SM{} and Ref.~\cite{theory}.

\section{Results}
\subsection{Optical properties of the microcavity}

\begin{figure}[h!tbp]
    \centering
    \includegraphics[width=\linewidth]{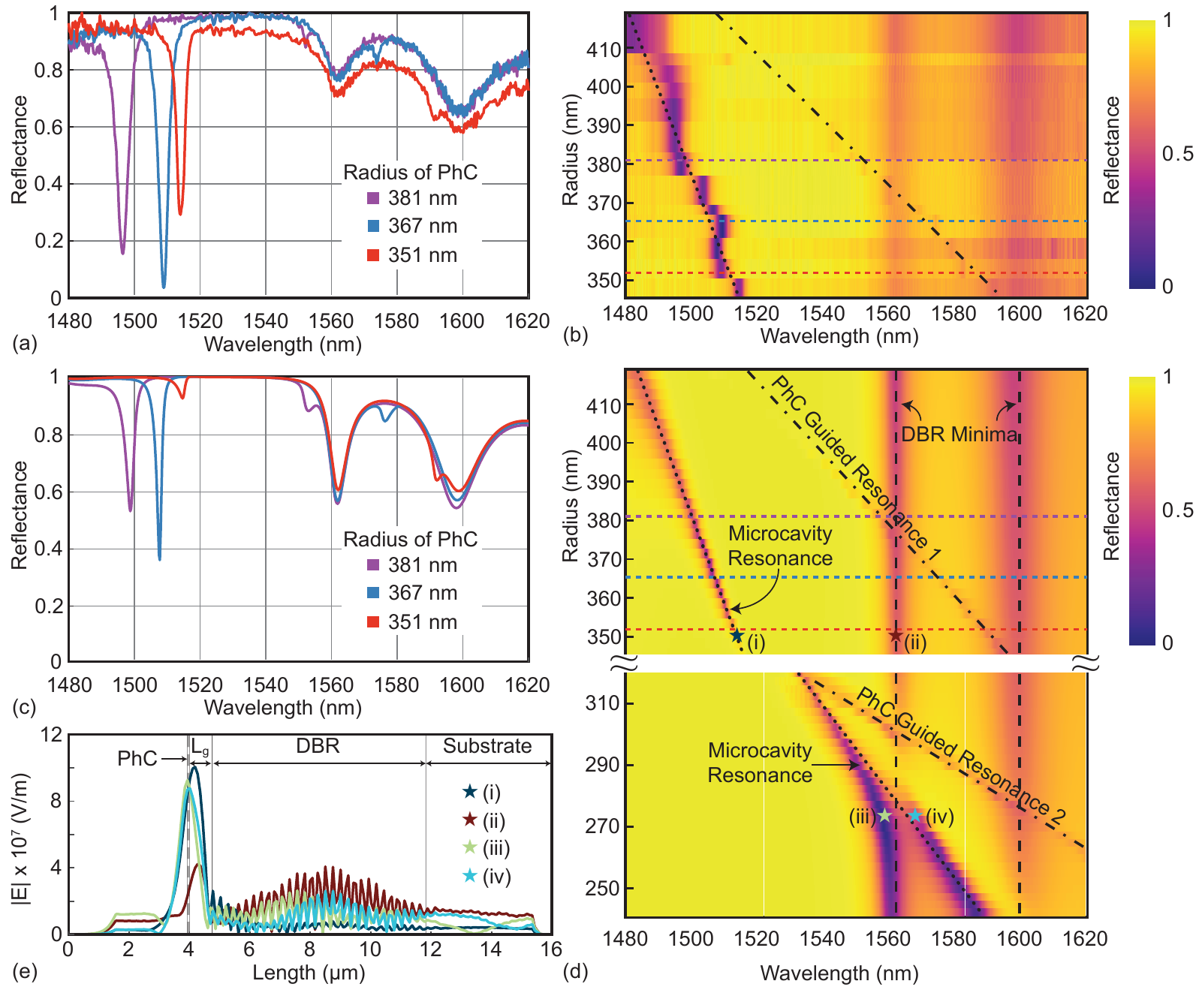}
    \caption{Characterization of the optical properties of the microcavity. (a) Reflectance spectra of a microcavity with PhC parameters: $a_\PhC = 1081\,$nm, $r_\PhC =$ 351, 367 and 381\,nm. (b) Reflectance map of the microcavity for varying PhC radii. (c) Simulation of reflectance spectra of a microcavity with parameters as in (a). (d) Simulation of a reflectance map of the microcavity for various PhC radii. The dashed, dotted, and dotted-dashed lines represent the DBR minima, microcavity resonance, and parasitic PhC-guided resonances, respectively. (e) The electric field distribution $|E|$ in the microcavity and DBR at the points marked in (d).}
    \label{fig:opticalchar}
\end{figure}

The optomechanical microcavity is formed by the PhC and DBR mirrors. While the stop-band of the DBR is determined by the heterostructure growth, the reflectance of the membrane can be engineered by the pattern of the PhC \cite{fan2002,makles_2d_2015}. We use this flexibility to demonstrate the tunability of the microcavity resonance wavelength by only varying the radius of the PhC holes, without having to adjust the geometric length of the microcavity, i.e., the gap between the PhC and DBR mirrors. A change of the PhC parameters changes the effective length of the microcavity, $L_\text{eff}$, which is given by the geometric cavity length, $L_\text{g}$, and an additional length originating from the phase acquired by the light in the PhC and DBR mirrors \cite{kim2007dbr}. We simulate the PhC membrane's reflectance using rigorous coupled wave analysis (RCWA) \cite{liu2012s4,kini2020suspended} to obtain the PhC parameters, i.e., PhC hole radius, $r_\PhC$, and PhC lattice constant, $a_\PhC$, such that the PhC yields its highest reflectance between 1480\,nm to 1520\,nm, which coincides with the reflectance band of the DBR.

\fref{fig:opticalchar}(a) shows the measured reflectance spectra of three fabricated microcavities with varied PhC hole radius and $a_\PhC =  1081$\,nm. We observe three distinct features in these spectra, which we exemplarily discuss for the microcavity with a PhC hole radius of 367\,nm. The microcavity resonance at 1508\,nm appears as a pronounced minimum in the reflectance spectrum. It exhibits an optical quality factor of about $3\cdot 10^2$. Its lineshape is slightly asymmetric due to the wavelength-dependent reflectance of the PhC slab \cite{Naesby2018}. Further, we observe a shallow reflectance dip at 1573\,nm. This dip originates from light coupling to a parasitic guided resonance of the PhC membrane due to the finite waist of the laser beam \cite{kini2020suspended,moura_centimeter-scale_2018}. Finally, two local minima at 1560\,nm and 1600\,nm are the first two transmission minima of the DBR mirror as confirmed by an independent measurement on a bare DBR (see \SM). 

In \fref{fig:opticalchar}(c) we present simulation results of the optical response for microcavities with parameters as in \fref{fig:opticalchar}(a). We obtain a good agreement between measurement and simulation for the features discussed above. However, we also observe  discrepancies, such as a deviation in the depth of the microcavity resonance and DBR minima. We attribute these differences to assumptions made in the RCWA simulation and approximations we make to simplify our model. The RCWA simulations assume a PhC that is in-plane infinitely periodic, while in the experiment, its size is given by the suspended area of the PhC membrane. Such finite-sized PhC structure reduces its absolute reflectance \cite{Jacob2000_finitephc}. This reduced reflectance of the PhC membrane does not alter the cavity resonance wavelength but affects the reflectivity mismatch between the PhC and DBR mirrors, which determines the microcavity dip depth. Another difference stems from fabrication imperfections of the PhC pattern, where the PhC parameters slightly vary over the exposed area and the profile of the sidewall of the PhC hole slightly deviates from being perfectly vertical. The variation of the PhC hole radii on the same device will reduce the absolute reflectance and, thus, add another contribution to a reflectivity mismatch between the PhC and DBR mirror affecting the depth of the microcavity resonance. The deviation from the ideal vertical hole profile leads to an increase in the loss of the PhC guided resonance due to coupling between TE and TM modes inside the PhC slab \cite{Tanaka2003,Khankhoje_2010}, resulting in a change of the PhC reflectance. Finally, we attribute the discrepancy between the simulated and measured depth of the DBR minima to the non-uniform thickness of the grown GaAs and (Al,Ga)As layers of the DBR, which we assume to be perfectly uniform in the simulation. The non-uniform layer thickness will reduce the overall reflectivity of the DBR reflection band and will also result in an increase of the reflectance at the DBR minima outside of the reflection band (see \SM).

To showcase the precise control over the microcavity resonance wavelength, we fabricated 14 devices with PhC hole radius varied from 351\,nm to 420\,nm and $a_\PhC= 1081$\,nm on a single chip, whose measured and simulated reflectance spectra are shown in \fref{fig:opticalchar}(b) and \fref{fig:opticalchar}(d), respectively. We observe a clear trend of the microcavity resonance wavelength that increases at a rate of 0.49\,nm per 1\,nm decrease in PhC hole radius with a similar value of 0.45\,nm per 1\,nm for the simulated case. Note that we attribute the slight fluctuation in the microcavity position [see Fig.~3(b)] around the expected behavior [see Fig.~3(d)] to a fabrication-related uncertainty in the exact geometric length of the microcavity. We further observe that the wavelength of the parasitic guided resonance shifts to larger wavelengths with a decrease in PhC hole radius, as observed in Refs.~\cite{kini2020suspended,moura_centimeter-scale_2018,Singh2013,Bernard2016}, and captured in the simulation [see Fig.~3(d)].

Up until $r_\PhC=310$\,nm, the position of the two DBR minima is independent of the PhC hole radius. Upon further reduction of the PhC hole radius, the microcavity resonance approaches the first transmission minimum of the DBR mirror and an avoided crossing between the DBR minimum and the microcavity is observed. In general, the DBR reflection minima are eigenmodes of the DBR structure and have applications, for example, in distributed feedback lasers \cite{ghafouri-shiraz}. These reflection minima are a result of destructive interference of the electric field in the DBR layers such that light is perfectly transmitted. In the presence of the microcavity field, the interference effect is disturbed, leading to the formation of new eigenmodes. To examine these new eigenmodes, we simulated the electric field distribution for a PhC hole radius of 275\,nm at the two branches of the avoided crossing, see \fref{fig:opticalchar}(e). Similarly, the electric field distribution of the microcavity mode and the eigenmode of the DBR far away from the avoided crossing for a PhC hole radius of 350\,nm is simulated for comparison. At the PhC hole radius of 350\,nm and at the microcavity resonance (i), the electric field is predominantly concentrated in the PhC membrane and vacuum gap. The DBR eigenmode at a PhC hole radius of 350\,nm (ii) is mostly contained in the DBR. At the avoided crossing (iii-iv), we observe that the microcavity and DBR eigenmode couple, and the energy is distributed between (i) and (ii) modes. This mode coupling is very similar to a system analyzed in Ref.~\cite{MahlerDFB2010}, which studied the coupling between an out-of-plane microcavity and an in-plane distributed feedback laser. The observed avoided crossing could have interesting applications in optomechanics. For example, one could functionalize the DBR layers by using lasing-active materials such that the optomechanical microcavity could be internally optically pumped \cite{yang_laser_2015,yu_active_2022}.

\subsection{Mechanical properties of the suspended PhC membrane}

\begin{figure}[t!bhp]
\centering\includegraphics{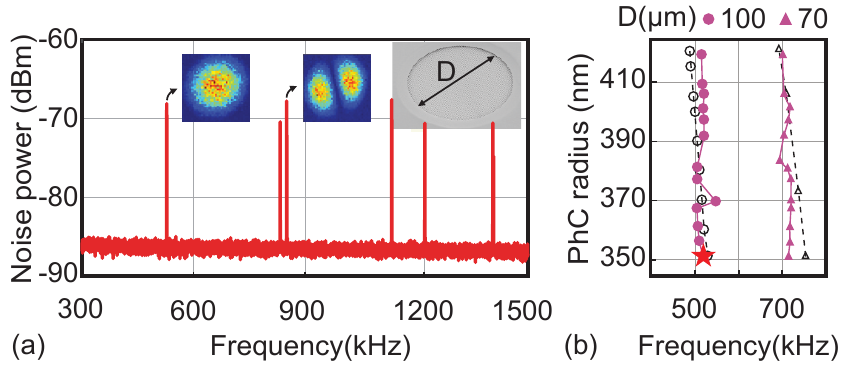}
    \caption{Characterization of the mechanical eigenfrequencies of the suspended PhC membrane. (a) Displacement noise power spectrum (NPS) of the thermally driven mechanical motion of a membrane of diameter 100\,$\mu$m (SEM image in the top right corner). The insets show the recorded mechanical mode tomography of the first two mechanical eigenmodes. (b) Measured frequencies of the fundamental mode of PhC membranes of two different diameters, where $r_\PhC$ is varied. The triangles (circles) represent PhC mirrors with a diameter of 70\,$\mu$m (100\,$\mu$m). The red star marks the device whose NPS is presented in (a). The black triangles (circles) show the frequencies of the fundamental mode of 70\,$\mu$m (100\,$\mu$m) PhC mirrors simulated via FEM. Lines are a guide to the eye.}
    \label{fig:mech}
\end{figure}

We now turn to the characterization of the mechanical properties of the suspended PhC membrane. \fref{fig:mech}(a) shows a representative displacement noise power spectrum of a circular PhC membrane with a diameter of 100\,$\mu$m. The fundamental mechanical mode is observed at 528\,kHz and the first higher order mode at 834\,kHz. We can attribute the modes through measurements of the spatially-dependent mode displacement recorded via mechanical mode tomography \cite{kini2020suspended}, shown in the insets of \fref{fig:mech}(a). The observed mode shapes reflect the expected behavior of a circular membrane and, thus, prove a complete and clean release of the PhC membrane on top of the 750\,nm air gap over the DBR mirror. When decreasing the diameter of the released PhC membrane, we observe an increase in the mechanical resonance frequency, see \fref{fig:mech}(b). While the variation of the PhC hole radius has a large impact on the optical resonance wavelength of the microcavity, we observe in \fref{fig:mech}(b) that the mechanical frequency of the fundamental mode only slightly decreases with an increase in PhC hole radius. This behavior is anticipated as the mechanical frequency is proportional to the ratio of the stiffness and mass of the PhC membrane, which both increase with a decrease in PhC hole radius, thus, leading to only a slight change in mechanical frequency. To confirm this observation, we performed FEM simulations of the PhC membranes [dashed line in \fref{fig:mech}(b)], whose results reflect the experimental measurements when assuming an initial tensile stress of 45\,MPa in the device layer\cite{kini2020suspended} (see \SM{}).

We also measured the mechanical quality factor $Q_\m$ via ringdown measurements and obtain values of $\leq 3 \cdot 10^{4}$. Note that $Q_\m$ can be vastly increased by substituting GaAs with tensile-strained InGaP \cite{cole2014,Buckle2018,manjeshwarMicromechanicalHighQTrampoline2022}, which is fully compatible with the presented monolithic (Al,Ga)As heterostructure growth and microfabrication process and allows applying strain engineering techniques to drastically increase $Q_\m$ \cite{Norte2016,tsaturyan_ultracoherent_2017,ghadimi_elastic_2018}, e.g., via the use of trampoline-shaped resonators\cite{Norte2016,Reinhardt2016,Romero_SiC_2020,manjeshwarMicromechanicalHighQTrampoline2022}. 

\subsection{Microcavity optomechanics}

\begin{figure*}[t!bhp]
    \centering\includegraphics{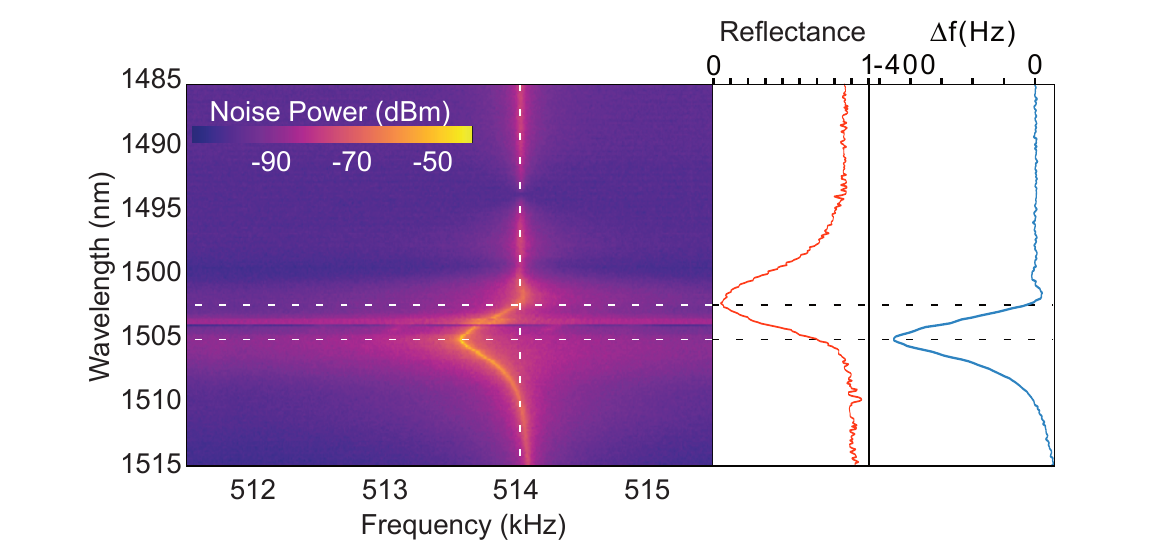}
    \caption{Microcavity optomechanics. The thermally driven noise power spectrum of the fundamental mode of the suspended PhC membrane shows a pronounced frequency shift for laser detunings that are red-detuned with respect to the microcavity resonance. The panel on the right shows the optical reflectance of the microcavity and the extracted mechanical frequency shift of the mechanical mode. The horizontal lines indicate the optical cavity resonance and the wavelength of the maximal mechanical frequency shift.}
    \label{fig:optomechanics_overview}
\end{figure*}

The mechanical properties of the suspended PhC mirror can be altered by the interaction of its motion with the microcavity field. \fref{fig:optomechanics_overview} shows the displacement noise power spectrum (NPS) of the fundamental mode of the suspended PhC mirror in dependence on the wavelength of the incident laser light. We clearly observe a non-symmetric mechanical frequency shift with respect to the microcavity resonance, which is predominantly to smaller frequencies. These features are also present in other devices, where we vary the PhC hole radius. In the presented device, the maximal shift of about $-400$\,Hz occurs at a pump wavelength of 1505\,nm, which is 3\,nm red-detuned with respect to the microcavity resonance. Such characteristics are not expected from canonical optomechanics \cite{Aspelmeyer2014Dec}. When increasing the input power to the microcavity, we observe that the mechanical frequency shift increases accordingly (see \SM{}). We verified that the optical reflectivity of the microcavity is not dependent on the optical input power (see \SM{}), which allows us to exclude nonlinear optical effects \cite{combrieGaAsPhotonicCrystal2008,parrainOriginOpticalLosses2015}.

\begin{figure*}[t!bhp]
    \centering\includegraphics{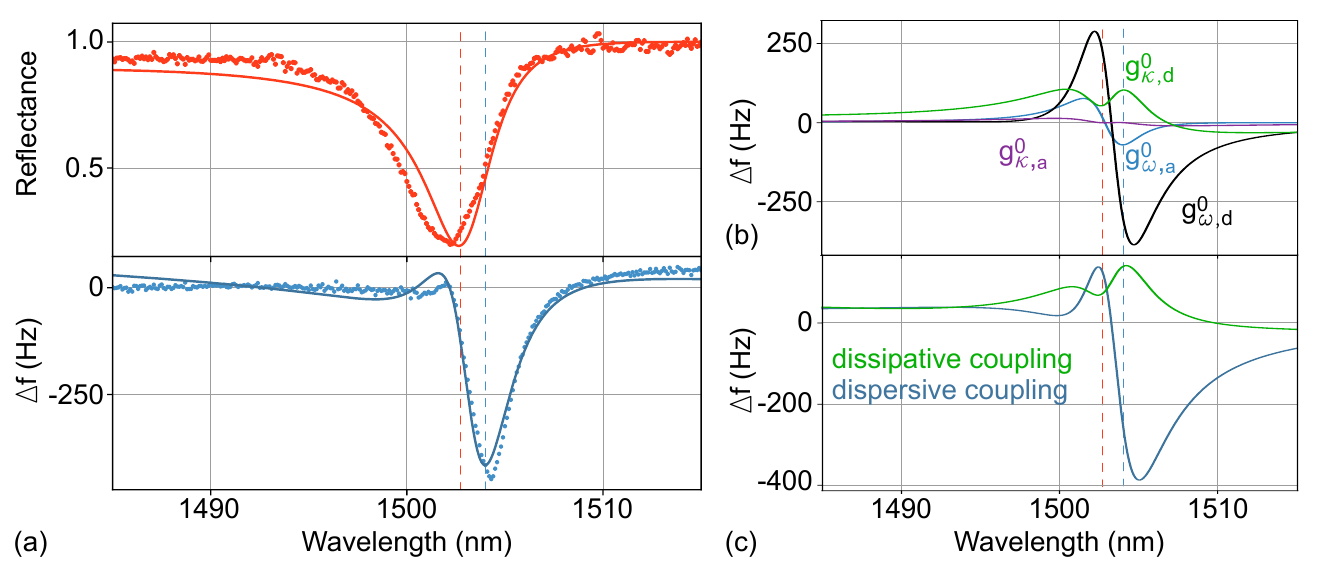}
    \caption{(a) Microcavity reflectance (top panel) and frequency shift (bottom panel). Theoretical values are presented in solid line and experiment in dots.
    For this plot $\gamma_d/2\pi = 3.8\,$THz, $\lambda_d = 1473$\,nm, $\Om/2\pi = 514$\,kHz, $\lambda_a/2\pi=1473.3$\,nm, $\kL/2\pi=2.12$\,THz,  $\kR/2\pi=366$\,GHz, $\Lambda/2\pi = 4.09$\,THz. Coupling strengths $\gka^0/2\pi = 775$\,kHz, $\gkd^0/2\pi = 3.21$\,MHz, $\gwa^0/2\pi = 845$\,kHz, $\gwd^0/2\pi = -1.82$\,MHz.
    (b,c) Comparison of coupling strength contributions to the frequency shift: (b) contribution of individual coupling (no cross-terms are taken into account), (c) only dissipative ($\gka^0$ and $\gkd^0$) or dispersive ($\gwa^0$ and $\gwd^0$) coupling are considered. The vertical dashed lines denote the microcavity resonance (red) and wavelength of maximal mechanical frequency shift (blue).
    }
    \label{fig:optomechanics_fit}
\end{figure*}

As outlined in the theory section, we account for dispersive and dissipative couplings of the mechanical mode of the suspended PhC to the Fabry-Pérot-type mode of the microcavity and to the guided mode of the PhC. Fig.\ref{fig:optomechanics_fit}(a) shows a combined fit of our model to the data of the optical reflectivity and of the mechanical frequency shift. We observe that we can reproduce the asymmetric optical lineshape and the off-resonant frequency shift of the mechanics. Fig.\ref{fig:optomechanics_fit}(b) shows the mechanical frequency shift, when we switch off all but one coupling. We observe that the shifts from individual couplings alone do not add up to the full mechanical frequency shift from panel (a) because of the presence of cross-terms involving different kinds of optomechanical couplings  (see \SM) and none of the individual couplings can capture well the observed frequency shift. Fig.\ref{fig:optomechanics_fit}(c) presents the mechanical frequency shift, when we switch off either the dissipative or the dispersive couplings, but leave the other ones on. We observe that the dispersive coupling captures the experimental behavior to a great extent, while the dissipative coupling yields a small correction. The major contribution comes from $\gwd^0$ while the one from $\gka^0$ is negligible.

Importantly, we find that the optical coupling $\Lambda$ between the cavity mode $\hat{a}$ and PhC mode $\hat{d}$ is in the strong coupling regime, $\Lambda > \kappa, \gd$, while the optomechanical couplings $g_i^0$ can be treated as a perturbation. We attribute this strong optical coupling to the hybridization of the PhC-guided resonance and the Fabry-Pérot-type mode [see the electric field in \fref{fig:sem_1}(d)]. Therefore, using \eref{eq:eigenfreq}, we determine that the effective resonances of the microcavity are at wavelengths $\lambda_+ = 1444$\,nm and $\lambda_- = 1503$\,nm and the corresponding loss rates are $\kappa_+/2\pi = 6.01$\,THz and $\kappa_-/2\pi = 277$\,GHz, respectively. The values of $\lambda_-$ and $\kappa_-$ are in good agreement with the experimental results shown in \fref{fig:optomechanics_fit}(a) and are associated with the effective optomechanical couplings $g_{\omega, -}^0/2\pi = -496$\,kHz and $g_{\kappa,-}^0/2\pi = 33.3$\,kHz, see \eref{eq:OM_eff_couplings}.

Importantly, we can connect the fitted parameters from \fref{fig:optomechanics_fit} to parameters that are accessible from the experiment or simulations. We outline this procedure in the \SM{}. We find reasonable experimental values for most of these parameters. An important quantity is the dominating coupling strength, $g_{\omega, -}^0/2\pi$, which we determine as a moving boundary coupling of the mechanical motion of the PhC to the microcavity optical field by FEM simulations \cite{johnsonPerturbationTheoryMaxwell2002, chanOptimizedOptomechanicalCrystal2012} (see supplementary material, Table S2). We obtain a simulated value that is wavelength-dependent and that is maximally $\sim - 2\pi\cdot240$\,kHz. This value is only about a factor of two smaller than the value we obtain from the fit. 

To further characterize the microcavity, a desired feature would be to tune the vacuum gap between the PhC and DBR mirrors. This gap impacts the optomechanical coupling strength and character as it will modify the coupling strength between the PhC guided mode and the Fabry-Pérot mode. This gap cannot be tuned in-situ in our monolithic structure but could be realized by the growth of wafers with varying sacrificial layer thickness. 

\section{Conclusions and outlook}

We demonstrated an on-chip free-space optomechanical microcavity of sub-wavelength length monolithically fabricated from an (Al,Ga)As heterostructure. We showed the tunability of the cavity resonance wavelength by a simple change of the radius of the PhC holes. The interplay between the Fabry-Pérot mode and the guided resonance of the suspended PhC resulted in a modified optomechanical response, which we modeled by accounting for dispersive and dissipative optomechanical couplings. Future studies with variable gaps are required to separate these couplings clearly from each other.

The presented optomechanical microcavity is amenable for optomechanical state preparation protocols, such as efficient cooling \cite{Cernotik2019Jun} or squeezed mechanical state generation \cite{kusturaMechanicalSqueezingUnstable2022}, the latter exploiting the ultrastrong coupling regime, which our microcavity accesses. Our approach allows for a smooth integration of multiple suspended PhC slabs and, thus, the realization of multi-element optomechanics \cite{Xuereb2013} on chip, which has been proposed to achieve the elusive single-photon strong coupling regime \cite{rablPhotonBlockadeEffect2011,nunnenkampSinglePhotonOptomechanics2011}. Additionally, our approach would enable the realization of arrays of optomechanical microcavities on a single chip with applications in sensing and optical networks \cite{wachter_silicon_2019}.  Finally, by use of the concept of bound states in the continuum realized with multiple PhC slabs \cite{fitzgerald2021cavity}, it is possible to drastically reduce optical loss,  which is the current roadblock in achieving strong optomechanical coupling physics.

\begin{acknowledgments}
We acknowledge the group of Victor Torres Company for their support with a tunable telecom laser, and Joachim Ciers for fruitful discussions. This work was supported by Chalmers' Area of Advance Nano, the Knut and Alice Wallenberg Foundation through a Wallenberg Academy Fellowship (W.W.) and the Wallenberg Center for Quantum Technology (WACQT, A.C.), the Swedish Research Council (VR projects No.~2018-05061 and 2019-04946), and the QuantERA project C'MON-QSENS! H.P. acknowledges funding by the European Union under the project MSCA-PF-2022-OCOMM. Samples were fabricated in the Myfab Nanofabrication Laboratory at Chalmers and analyzed in the Chalmers Materials Analysis Laboratory. Simulations were performed on resources provided by the National Academic Infrastructure for Supercomputing in Sweden (NAISS) and the Swedish National Infrastructure for Computing (SNIC) at Tetralith partially funded by the Swedish Research Council through grant agreements no. 2022-06725 and no. 2018-05973.
\end{acknowledgments}

\clearpage

\addcontentsline{toc}{section}{Supplementary Material}
\section*{Supplementary Material}

\beginsupplement


\section{Theory}

\subsection{Coupled-mode theory applied to optomechanical system}\label{sec:CM}
This section explains how the microcavity reflectance and mechanical frequency shift can be derived from the coupled-mode model presented in the theory part of the main text and analyzed in detail in Ref.~\cite{theory}.

\subsubsection{Langevin equations}

\begin{figure*}[t!bhp]
    \centering
    \includegraphics{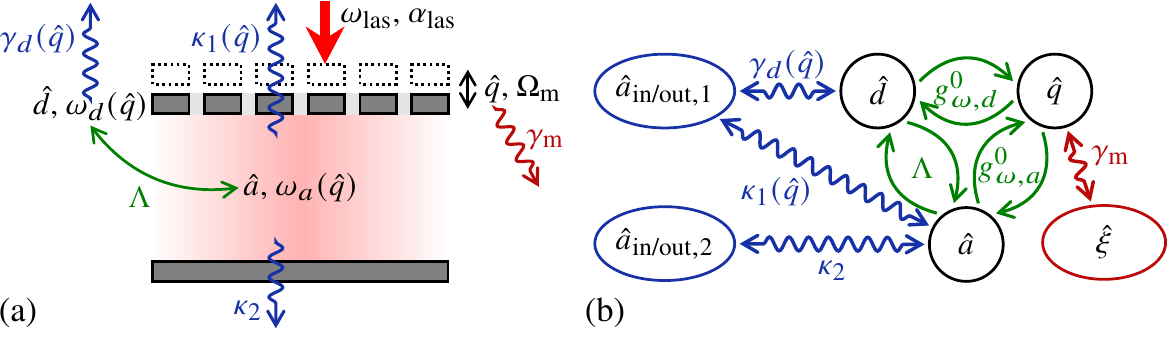}
    \caption{Coupled-mode model of the PhC-DBR microcavity: (a) schematic of the microcavity and (b) coupled-mode picture. The green arrows represent Hamiltonian couplings, the wavy arrows correspond to dissipative couplings to photonic (in blue) and phononic (in red) environments.}
    \label{fig:omschematic}
\end{figure*}

As explained in the main text, the PhC-DBR microcavity, depicted in \fref{fig:omschematic}, is modeled with three coupled modes: two optical modes and one mechanical mode.
Due to the dispersive mechanical couplings, we have, at the first order in the mechanical displacement, $\oma(\hat{q}) \simeq \oma - \sqrt{2} \gwa^0 \hat{q}$ and  $\omd(\hat{q}) \simeq \omd - \sqrt{2} \gwd^0 \hat{q}$. The system's Hamiltonian therefore reads
\begin{widetext}
\begin{equation}
    \hat{H} =  \hbar\oma \hat{a}^\dagger \hat{a}+\hbar\omd \hat{d}^\dagger \hat{d} + \hbar\Lambda(\hat{a}^\dagger\hat{d} + \hat{a}\hat{d}^\dagger) +  \frac{\hbar\Om}{2}(\hat{q}^2 + \hat{p}^2) -\hbar\gwa^0\sqrt{2}\hat{a}^\dagger \hat{a}\hat{q}-\hbar\gwd^0\sqrt{2}\hat{d}^\dagger \hat{d}\hat{q}.
\end{equation}
\end{widetext}
The cavity mode is coupled to two photon baths through the top (PhC) and bottom (effective mirror, DBR) mirror with optical loss rates $\kL$ and $\kR$, respectively. The mirror mode is coupled to the top bath only, with loss rate $\gd$. At the first order in the mechanical displacement, the dissipative optomechanical effects give $\kL(\hat{q}) \simeq \kL + \sqrt{2} \gka^0 \hat{q}$ and $\gd(\hat{q}) \simeq \gd + \sqrt{2} \gkd^0 \hat{q}$. Then, the evolution of this system can be described with a set of Langevin equations~\cite{Cernotik2019Jun, Elste2009May, theory}:
\begin{widetext}
\begin{align}\label{eq:Langevin}
    \dot{\hat{a}}=\,& -(i\Delta_a + \kappa) \hat{a} + \sqrt{2}(i \gwa^0 - \gka^0)\hat{q}\hat{a}-\G \hat{d}- \sqrt{2}\gadp \hat{q}\hat{d}\\\nonumber
    &+ \left(\sqrt{2\kL} + \frac{\gka^0}{\sqrt{\kL}} \hat{q}\right)(\aL^{} + \ainL^{}) + \sqrt{2\kR}\ainR^{},\\\nonumber
    \dot{\hat{d}}=\,& -(i\Delta_d + \gd) \hat{d} +\sqrt{2}(i \gwd^0 - \gkd^0)\hat{q}\hat{d} -\G \hat{a} - \sqrt{2}\gadp\hat{q}\hat{a} + \left(\sqrt{2\gd} + \frac{\gkd^0}{\sqrt{\gd}} \hat{q}\right)(\aL^{} + \ainL^{}),\\\nonumber
    \dot{\hat{q}}=\,& \Om \hat{p},\\\nonumber
    \dot{\hat{p}}=\,& -\Om\hat{q} -\gam \hat{p}  + i\sqrt{2}\gadm(\hat{a}^\dagger\hat{d}- \hat{d}^\dagger\hat{a}) + \sqrt{2}\gwa^0  \hat{a}^\dagger \hat{a}  + \sqrt{2}\gwd^0  \hat{d}^\dagger \hat{d}  + \sqrt{\gam}\hat{\xi}\\\nonumber
    &- i\frac{\gka^0}{\sqrt{\kL}}(\hat{a}^\dagger (\aL^{} + \ainL^{}) - (\aL^* + \ainL^\dagger) \hat{a}) - i\frac{\gkd^0}{\sqrt{\gd}}(\hat{d}^\dagger (\aL^{} + \ainL^{}) - (\aL^* + \ainL^\dagger) \hat{d}).
\end{align}
\end{widetext}
The mechanical quality factor $Q_\m = \Om/\gam$ is large and the temperature $T$ of the mechanical environment is high, that is $\hbar\Om \ll \kB T$, therefore we make the white-noise approximation for the correlation function for the thermal noise of the mechanics
\begin{equation}
    \mean{\hat \xi(t)\hat \xi(t')}=(2\Nm+1) \delta (t-t'),\label{correlations_xi}
\end{equation}
where $\Nm = (\e^{\hbar\Om/\kB T_\m} - 1)^{-1}$ is the average number of phonons in the mechanical environment.
The laser driving the cavity is modeled as a classical drive of complex amplitude $\aL$ and treated as part of the top input field. The input laser power is given by 
$\Pin = \hbar\omL \abs{\aL}^2$.  The average number of phonons in the optical environment is negligible since $\hbar \oma \gg \kB T$, therefore, there is only one non-zero correlation function of the optical input noises:
\begin{equation}
    \mean{\ain^{}(t)  \ain^\dagger(t')}= \delta (t-t'), \label{correlations_a_in}
\end{equation}
with $\mu = \text{1,2}$.

\subsubsection{Mechanical frequency shift}

To obtain the mechanical frequency shift, we first linearize the Langevin equations, \eref{eq:Langevin}, by splitting operators into their average terms and fluctuation terms, $\hat{a} = \alpha + \da$, $\hat{q} = \bar{q} + \dq$, $\hat{p} = \bar{p} + \deltp$ and $\hat{d}=\delta + \deltd$, and keeping only the first order in the fluctuations. Taking the average of \eref{eq:Langevin} gives a nonlinear system of equations which, in our case, namely $\kappa, \gd \gg \Om, \gwa^0, \gwd^0, \gka^0, \gkd^0$ (see Table~\ref{tab:param}), has a unique steady-state ($\alpha$, $\delta$, $\bar{q}$, $\bar{p}$).
Due to the average mechanical displacement $\bar{q}$, the detunings and loss rates of the optical modes become $\bar{\Delta}_{a/d} = \Delta_{a/d} - \gw[a/\smash{d}]^0\sqrt{2}\bar{q}$, $\bar{\kappa}_\L = \kL + \gka^0 \sqrt{2} \bar{q}$ and $\bar{\gamma}_d = \gd + \gkd^0 \sqrt{2} \bar{q}$, but in our case, this shift is negligible. Therefore, the dynamics of the fluctuations is given by
\begin{widetext}
\begin{align}\label{Langevin lin eqs}
    \delta\dot{\hat{a}} =\,&-(i\Da + \kL + \kR) \da +i\sqrt{2}\tga\dq -\G\deltd  + \sqrt{2\kL}\ainL+ \sqrt{2\kR}\ainR,\\\nonumber
    \delta\dot{\hat{d}} =\,&-(i\Dd + \gd) \deltd +i\sqrt{2}\tgd\dq  -\G\da + \sqrt{2\gd}\ainL ,\\\nonumber
    \delta\dot{\hat{q}} =\,&\Om \deltp,\\\nonumber
    \delta\dot{\hat{p}}=\,& -\Om\dq -\gam \deltp+  \sum_{c=a,d}\sqrt{2} (\tgm \delta\hat{c}^\dagger +\tgm^*\delta\hat{c})\nonumber +C^*\ainL + C\ainL^\dagger+ \sqrt{\gam}\hat{\xi},
\end{align}
where we have defined the effective optomechanical couplings
\begin{align}\label{eq:effective couplings}
    \tga &=  \gwa^0\alpha +i \gka^0\left(\alpha  -\frac{\aL}{\sqrt{2\kL}}\right) + i\gadp\delta,&  \tgma &= \gwa^0\alpha - i\gka^0\frac{ \aL}{\sqrt{2\kL}}+ i\gadm\delta,\\
   \tgd &= \gwd^0\delta +i \gkd^0\left(\delta -\frac{\aL}{\sqrt{2\gd}}\right) + i\gadp\alpha,&  \tgmd &= \gwd^0\delta - i\gkd^0\frac{ \aL}{\sqrt{2\gd}}-i\gadm\alpha,\nonumber
\end{align}
\end{widetext}
and the coefficient $C = i\left(\gka^0\alpha/\sqrt{\kL}+\gkd^0\delta/\sqrt{\gd}\right)$.

Then, we solve this linear system of equations in the frequency domain, which allow us to determine the effective mechanical susceptibility \cite{Aspelmeyer2014Dec, theory},
$\chi_\m^{-1}[\omega] = \chi_{\m,0}^{-1}[\omega] + \chi_\text{opt}^{-1}[\omega]$,
where $\chi_{\m,0}[\omega] = \Om\left(\Om^2 - \omega^2 - i\omega\gam\right)^{-1}$ is the mechanical susceptibility of the bare resonator and $\chi_\text{opt}[\omega]$ the optical contribution to the susceptibility, which reads
\begin{align}
    \chi_\text{opt}^{-1}[\omega]= -2\sum_{c} \left(\tgm^*C_q^c[\omega] +\tgm C_q^c[-\omega]^*\right),\label{eq:Xopt}
\end{align}
with
\begin{align}\label{Cq_c}
    C^a_q[\omega] &= i\frac{\chi_d^{-1}[\omega]\tga - \G \tgd}{\chi_a^{-1}[\omega]\chi_d^{-1}[\omega] - \G^2},\\
    C^d_q[\omega] &=i\frac{\chi_a^{-1}[\omega]\tgd - \G \tga}{\chi_a^{-1}[\omega]\chi_d^{-1}[\omega] - \G^2}.
\end{align}
Finally, we put $\chi_\m[\omega]$ in the usual form for a harmonic oscillator~\cite{Aspelmeyer2014Dec}, identifying the frequency shift due to the optical spring effect 
\begin{equation}\label{eq:dOm}
    \delta\Om[\omega] = \frac{1}{2}\Re{(\chi_\text{opt}[\omega]^{-1})}.
\end{equation}
In the regime relevant for this experiment, the frequency dependence is negligible and we use the frequency shift $\delta\Om[\Om]$ to fit the experimental data and the $\Delta f$ plotted in Fig. 6 in the main text corresponds to $\delta\Om[\Om]/2\pi$. Note that it is not possible to split this mechanical frequency shift in independent contributions from individual optomechanical couplings. It is clear from Eqs.~\ref{eq:Xopt} and \ref{eq:effective couplings} that there are many cross-terms involving different couplings.

%
\subsubsection{Reflectance of the microcavity using input-output relation}

In the experiment, the reflectance $R$ of the microcavity (from the top) is measured. We can compute this quantity from the coupled-mode model with $R = 1 - \abs{t}^2$, where  $t = \mean{\aoutR}/\aL$ is the amplitude transmission coefficient of the cavity. $\aoutR$ is the output field through the bottom mirror, given by the input-output relation $\aoutR(t) =  \ainR(t) - \sqrt{2\kR}\hat{a}(t)$ \cite{Gardiner1985Jun, theory}, and the average input field through the top mirror is the laser field $\aL$ while $\mean{\ainL} = \mean{\ainR} = 0$. 
Therefore, we obtain
\begin{widetext}
\begin{equation}\label{eq:SM_TransCM}
    t(\omL) = - \sqrt{2\kR}\frac{\alpha}{\aL} = \frac{\sqrt{2\kR}\left[\sqrt{2\gd}\G - \sqrt{2\kL} (\gd + i(\omd - \omL))\right]}{[\kappa + i(\oma - \omL)][\gd + i(\omd - \omL)] - \G^2 },
\end{equation}
\end{widetext}
where, like in the previous section, we have neglected the correction of the optical detunings and loss rates due to the average mechanical displacement $\bar{q}$.

\subsection{Transfer matrix modeling of optical components}\label{sec:TMModeling}

The optical transmission and reflection of the PhC-DBR microcavity is modeled by using the transfer matrix method, see, e.g., Ref.~\cite{deutsch1995photonics}. We begin by independently modeling the PhC slab and the DBR in vacuum, followed by the combined system. For further analysis we simplify the system by making approximations to replace the DBR mirror with a thin scatterer of high reflectance. 

The optical transmission and reflection of a scatterer can be described by its polarizability $\zeta$ given by \cite{deutsch1995photonics}:
\begin{equation}
    \zeta = \frac{r}{it}
    \label{eq:polarizability}
\end{equation}
where $r$ and $t$ is the reflection and transmission coefficient, respectively, and both are complex numbers and fulfill energy conservation \cite{deutsch1995photonics}. Alternatively, one can also express $r$ and $t$ in terms of $\zeta$ as follows:

\begin{align}
    r = \frac{i\zeta}{1 - i\zeta}\label{eq:r_w_zeta}\\
    t = \frac{1}{1 - i\zeta}.\label{eq:t_w_zeta}
\end{align}

Following conditions must be fulfilled such that $r$ and $t$ fulfill energy conservation \cite{deutsch1995photonics}:
\begin{align}
    |r|^2+|t|^2 & =1\nonumber\\
    rt^*+r^*t & =0 \label{eq:cond1}
\end{align}

 such that one can write 
\begin{align}
    r & = i\sin{\phi}e^{i\phi}\\
    t & = \cos{\phi}e^{i\phi},
\end{align}
with phase $\phi$. Using \eref{eq:r_w_zeta} and \eref{eq:t_w_zeta}, then it also holds that
\begin{align}
    t & =1+r. \label{eq:cond2}
\end{align}

The transfer matrix for a scatterer is then \cite{deutsch1995photonics}:
\begin{equation}
M_{r,t} = \frac{1}{t} 
    \begin{bmatrix}
        1  & -r \\
        r & t^2 - r^2 
    \end{bmatrix}.
    \label{mat:PhC_rt}
\end{equation}

The transfer matrix can be expressed in terms of the polarizability by replacing $r$ and $t$ with \eref{eq:r_w_zeta} and \eref{eq:t_w_zeta}:

\begin{equation}
M_{\zeta} =
    \begin{bmatrix}
        1 - i\zeta  & -i\zeta \\
        i\zeta & 1 + i\zeta
    \end{bmatrix}.
    \label{mat:PhC_zeta}
\end{equation}

The transfer matrix describing scattering at the interface between two media ($1\rightarrow2$) is given by \cite{a_teich_2019}

\begin{equation}
M_{\text{int} \,1\rightarrow 2}= \frac{1}{t_{1\rightarrow2}}
    \begin{bmatrix}
        1  & r_{1\rightarrow2} \\
        r_{1\rightarrow2} & 1
    \end{bmatrix},
    \label{mat:int1}
\end{equation}
with the Fresnel transmission and reflection coefficients for TE-polarization at normal incidence given as 
$r_{1\rightarrow2} = \frac{n_1 - n_2}{n_1 + n_2}$ and $t_{1\rightarrow2} = \frac{2 n_1}{n_1 + n_2}$ \cite{jackson1999classical}.

The transfer matrix describing propagation in a material with thickness $d$ and refractive index $n$ is given by 
\begin{equation}
M_{\text{prop}} = 
    \begin{bmatrix}
        e^{iknd}  & 0 \\
        0 & e^{-iknd}
    \end{bmatrix},
    \label{mat:int2}
\end{equation}
where $k=2\pi/\lambda$ is the wavevector and $\lambda$ the wavelength.

To obtain the transfer matrix $M_{\text{tot}}$ with elements $M_{\text{tot}}(i,j)$ of a complete optical system made up of optical components $1, 2, 3, ...$, one simply multiplies the individual transfer matrices:
\begin{equation}
    M_{\text{tot}} = M_1M_2M_3 \cdots
\end{equation}

The reflection coefficient $r_{\text{tot}}$ and transmission coefficient $t_{\text{tot}}$ of the complete system then can be found as
\begin{eqnarray}
    r_{\text{tot}} & = & \frac{M_{\text{tot}}(2,1)}{M_{\text{tot}}(1,1)},\\
    t_{\text{tot}} & = & \frac{1}{M_{\text{tot}}(1,1)}.
\end{eqnarray}

\subsubsection{PhC in vacuum}
\label{sec:TMM}

We consider the PhC membrane to be an infinitely thin lossless scatterer whose optical properties can be described with polarizability $\zeta_1(\omega)$ \cite{fan2003temporal,newsom2020optimal}. This approximation is valid for membranes where the thickness is small in comparison to the wavelength of light, thus we can ignore the phase shifts due to the propagation inside the dielectric \cite{nair2016cavity}.

The reflection and transmission coefficients of a PhC slab with a single Fano resonance is given by  \cite{fan2003temporal}
\begin{align}
    r_1(\omega) &= \frac{- s\gamma_d t_d(\omega) - i\Delta_d r_d(\omega)}{-i\Delta_d + \gamma_d} \label{eq:Fano_r_f}
    \\
    t_1(\omega) &= \frac{- s\gamma_d r_d(\omega) -i\Delta_d t_d(\omega)}{-i\Delta_d+\gamma_d}, \label{eq:Fano_t_f}
\end{align}
where $s =1(-1)$ for even (odd) mode with respect to the mirror plane in the middle of the slab, $\Delta_d = \omd - \omega$, $\omd$ is the Fano resonance frequency, $\gd$ is the linewidth of the Fano resonance, $t_d$ and $r_d$ are the direct transmission and reflection coefficients.

By substituting  \eref{eq:Fano_r_f} and \eref{eq:Fano_t_f} into \eref{eq:polarizability}, we obtain the frequency-dependent polarizability of a PhC slab $\zeta_{1}(\omega)$ : 

\begin{equation}\label{eq:Fano_PhC}
    \zeta_1(\omega) 
= {\zeta_0} \frac{s\gamma_d + \Delta_d/\zeta_0}{s\gamma_d -\Delta_d \zeta_0}
,
\end{equation}
where $\zeta_0 =\zeta_1(\omd) = \frac{t_d}{i r_d}$. Note that $r_d$ and $t_d$ themselves must fulfill  conditions \eref{eq:cond1} and \eref{eq:cond2}, such that $r_1(\omega)$ and $t_1(\omega)$ fulfill these conditions as well and, thus, are valid descriptions of the PhC slab as a scatterer. Hence, it is the easiest to write them as  $r_d(\omega)  = i\sin{\phi_d(\omega)}e^{i\phi_d(\omega)}$ and $t_d(\omega)  = \cos{\phi_d(\omega)}e^{i\phi_d(\omega)}$ with a phase $\phi_d(\omega)$ that depends on frequency.

\subsubsection{PhC - DBR cavity}

To simplify the modeling of the combined PhC-DBR system, we replace the DBR with a thin scatterer of polarizability $\zeta_2$. This thin scatterer approach models only the high reflectance region in the stop-band of the DBR, which is sufficient for the following analysis. The phase acquired from the DBR upon reflection is accounted for by an additional phase introduced in the free-space propagation matrix, so we have $\Phi = k L_g + \phi_{dbr}$, with $L_g$ the length of the vacuum gap. We, thus, account for an effective cavity length change.

The total optical system is then given by
\begin{widetext}
\begin{align}
M & =
\begin{bmatrix}
    1-i\zeta_1(\omega)  & -i\zeta_1(\omega) \\
    i\zeta_1(\omega) & 1 + i\zeta_1(\omega)
\end{bmatrix}
\begin{bmatrix}
    e^{i\Phi}  & 0 \\
    0 & e^{-i\Phi}
\end{bmatrix}
\begin{bmatrix}
    1-i\zeta_2  & -i\zeta_2 \\
    i\zeta_2 & 1 + i\zeta_2 
\end{bmatrix}\\
&= e^{i\Phi}
    \begin{bmatrix}
        (1 - i\zeta_1(\omega))(1 - i\zeta_2) + \zeta_2\zeta_1(\omega) e^{-i2\Phi} &  - i\zeta_2 (1 - i\zeta_1(\omega)) - i\zeta_1(\omega)(1 + i \zeta_2) e^{-i2\Phi} \\
        i\zeta_1(\omega) (1 - i\zeta_2) + i \zeta_2(1 + i\zeta_1(\omega))e^{-i2\Phi} & \zeta_1(\omega)\zeta_2 + (1 + i\zeta_2)(1 + i\zeta_1(\omega))e^{-i2\Phi}
    \end{bmatrix}
\end{align}
\end{widetext}
                          
The reflection coefficient of the complete system is then:
\begin{align}\label{eq:SM_ReflecTM}
    r_{1-2}(\omega) &= \frac{M(2,1)}{M(1,1)} \\
    &= \frac{i\zeta_1(\omega) (1 - i\zeta_2) + i \zeta_2(1 + i\zeta_1(\omega))e^{-i2\Phi} }{(1 - i\zeta_1(\omega))(1 - i\zeta_2) + \zeta_2\zeta_1(\omega) e^{-i2\Phi}},
\end{align}

and its transmission coefficient:
\begin{align}
    t_{1-2}(\omega) &= \frac{1}{M(1,1) } \\
    &= \frac{1}{(1 - i\zeta_1(\omega))(1 - i\zeta_2) e^{i\Phi} + \zeta_2\zeta_1(\omega) e^{-i\Phi}}
    \label{eq:SM_TransTM}.
\end{align}

\subsection{Parameters for the coupled-mode theory}

The parameters from the coupled-mode model from \sref{sec:CM} cannot be straightforwardly linked to experimentally measurable parameters. Indeed, in the experiment, we only have access to the effective optical properties of the microcavity and, therefore cannot measure separately the resonance frequencies and loss rates of the Fabry-Pérot cavity mode and mirror mode from the model. However, we can estimate the mirror mode parameters $\omd$, $\gd$ by fitting RCWA simulations with the transfer matrix approach from \sref{sec:TMModeling} and then obtain the remaining parameters, $\oma$, $\kL$, $\kR$, $\Lambda$, and the optomechanical couplings, by simultaneously fitting the experimental microcavity reflectance and mechanical frequency shift using Eqs.~\ref{eq:SM_TransCM} and \ref{eq:dOm}.

\subsubsection{Bridging transfer matrix and coupled-mode theory parameters}
Like in Ref.~\cite{Cernotik2019Jun}, we compare the transmission coefficients obtained by the transfer matrix model, \eref{eq:SM_TransTM}, and by the coupled mode theory, \eref{eq:SM_TransCM}, to connect the parameters used in these two methods. First, the linewidth $\gd$ and resonance frequency $\omd$ used in the transfer matrix formalism are the same in the coupled-mode theory. Indeed, in the limit where the Fano resonance becomes Lorentzian, it has the same characteristics as the bare mirror mode described by the equation $\dot{\hat{d}} = -(i\omd + \gd)\hat{d} + \sqrt{2\gd}\ainL$.
Second, in the transfer matrix model, the transmission of the microcavity becomes zero when $\zeta_\L(\omega)$ tends to infinity at frequency $\omega_0 = \omd - \gd/\zeta_0$. By enforcing $t(\omega_0) = 0$ in the coupled mode model, we get $\Lambda = \sqrt{\kL\gd}/\zeta_0$.

\subsubsection{Fitting procedure}
We use the following procedure to extract the values of the parameters:
\begin{enumerate}
    \item $r_\PhC,a_\PhC$ and $d_\PhC$: PhC parameters extracted from SEM, then compare the measured reflectance to RCWA simulations with the SEM parameters, see \sref{sec:microcavparams}.
    \item $\omd$, $\gd$, $\zeta_0$: Compute the reflectance of the PhC in vacuum with an RCWA simulation using found $r_\PhC,a_\PhC$ and $d_\PhC$ from step (1) and fit it with $\abs{r_1(\omega)}^2$, \eref{eq:Fano_r_f}, from the transfer matrix model, see \sref{sec:TMM}.
    \item $\oma$, $\kL$, $\kR$, $\Lambda$, $\gwa^0$, $\gwd^0$, $\gka^0$ and $\gkd^0$: Simultaneously fit the experimental microcavity reflectance and mechanical frequency shift with Eqs.~\ref{eq:SM_TransCM} and \ref{eq:dOm}, using $\omd$, $\gd$ from step (2) and the experimental values of $\Om$ and $\Pin$. We use $\Lambda = \sqrt{\kL\gd}/\zeta_0$ as a starting value but allow for adjustments of the value of $\Lambda$ (amounting to refitting for $\zeta_0$) since $\omd$, $\gd$, $\zeta_0$ we determined from simulations and not from experimental data.
\end{enumerate}
We tried to perform step (3) with less fitting parameters by neglecting some optomechanical couplings but apart from $\gka^0$ that has little influence here, all the other ones are required to reproduce the shape of the mechanical frequency shift.

In Table~\ref{tab:param}, we show the values of the parameters obtained in the experiment (SEM, TEM, reflectance and optical interferometer) and the ones allowing us to reproduce the experimental results with the coupled-mode theory (CM), RCWA simulations, the transfer matrix theory (TM). The global sign of the optomechanical couplings is not well determined as inverting all the signs gives the same frequency shift. We also note that their values are all quite high, in the ultrastrong-coupling regime, that is, larger than $\Om$. But these couplings tell us how the mechanical motion affects the frequencies of the theoretical Fabry-Pérot cavity and mirror modes, not how it affects the effective cavity resonance and linewidth, that are accessible in the experiment and in RCWA simulations. However, given that the two theoretical optical modes are strongly coupled, $\Lambda > \kappa, \gd$, the effective cavity resonance is well approximated by one of the eigenfrequencies of the coupled optical system, in this case $\tilde{\omega}_-$, see Eq.~7 in the main text. The corresponding resonance wavelength and loss rate are $\lambda_- = 1503$\,nm and $\kappa_-/2\pi = 277$\,GHz. They are associated with the effective couplings $g_{\omega, -}^0/2\pi = -496$\,kHz and $g_{\kappa,-}^0/2\pi = 33.3$\,kHz. 

Though still close to the ultrastrong regime, these values are more reasonable and $g_{\omega, -}^0$ is of the same order of magnitude as the moving boundary coupling of the mechanical motion of the PhC to the microcavity optical field obtained by FEM simulations \cite{johnsonPerturbationTheoryMaxwell2002, chanOptimizedOptomechanicalCrystal2012}, see Table~\ref{tab:fem_mb}. Furthermore, these simulations give an optical eigenmode at wavelength 1494 nm, relatively close to the $\lambda_-$ from the CM theory. The associated loss rate of 89 GHz is significantly lower than $\kappa_-$, but loss mechanisms like fabrication imperfections and diffraction losses are neglected in the simulations while they are captured by the fit of the experimental data.

\begin{table*}
    \begin{center}
        \begin{tabular}{ |c|c|c|c|c|c| } 
             \hline
             Parameters & Description & Experiment & CM & RCWA & TM \\ 
             \hline
             \multicolumn{6}{l}{Parameters of PhC}\\
             \hline
             $r_\PhC$ (nm)& hole radius & 355 $\pm$ 4 (SEM) & -  & 356  & - \\ 
             $a_\PhC$ (nm)& lattice constant & 1081 $\pm$ 4  (SEM) & -  & 1081 & - \\ 
             $d_\PhC$ (nm)&slab thickness & 75 $\pm$ 2.4  (SEM) & -  & 74 & - \\
             $\lambda_d$ (nm) & Fano resonance wavelength & - & 1473 & - & 1473\\ 
             $\gamma_d/2\pi$ (THz)& Fano linewidth & - & 3.8 & - & 3.8\\ 
             $\zeta_0$  & polarizability at resonance & - & 0.694 & - & 0.715\\ 
             \hline
             \multicolumn{6}{l}{Parameters of mechanical resonator}\\
             \hline
             $\Om/2\pi$ (kHz)& resonance frequency & 514 & 514 & - & -\\ 
             $\gam$ (Hz) & damping rate & 170  & 170  & -  &  - \\ 
             \hline
             \multicolumn{6}{l}{Parameters of microcavity}\\
             \hline
             $L_g$ (nm) & geometric length & 745 $\pm$ 4.5 (TEM) & - & 715 & 715\\ 
             $\lambda_\text{cav}$ (nm) & resonance wavelength & 1502.2 & 1502.6 & 1502& 1502.5\\ 
             $\kappa_\text{cav}/2\pi$ (GHz)& linewidth & 347 & 310 & 62.5 & -\\ 

             \hline
             \multicolumn{6}{l}{Parameters of Fabry-Pérot (FP) cavity mode}\\
             \hline
             $\lambda_a$ (nm)& resonance wavelength & - & 1473.3 & - & -\\ 
             $\kL/2\pi$ (THz)& coupling to environment 1 & - & 2.12 & - & - \\ 
             $\kR/2\pi$ (THz)& coupling to environment 2  & - & 0.366 & - & - \\ 
             $\Lambda/2\pi$ (THz)& PhC-FP cavity coupling & - & 4.09 & - & -\\
              \hline
             \multicolumn{6}{l}{Optomechanical couplings}\\
             \hline
             $\gwa^0/2\pi$ (kHz)& FP dispersive coupling & - & 845 & - & -\\ 
             $\gwd^0/2\pi$ (kHz)& PhC dispersive coupling & - & -1819 & - & -\\ 
             $\gka^0/2\pi$ (kHz)& FP dissipative coupling & - & 775 & - & -\\ 
             $\gkd^0/2\pi$ (kHz)& PhC dissipative coupling & - & 3213 & - & -\\ 
             $g_{\omega, -}^0/2\pi$ (kHz) & effective dispersive coupling &- & -496 & - & -\\  $g_{\kappa,-}^0/2\pi$ (kHz) & effective dissipative coupling &- & 33.3& - & -\\
             \hline
        \end{tabular}
        \caption{Parameters of the optomechanical microcavity used in the coupled mode theory modeling, in the transfer matrix modeling, and obtained from simulations or experiment.}
        \label{tab:param}
    \end{center}
\end{table*}

\begin{table*}
    \begin{center}
        \begin{tabular}{ |c|c|c| } 
             \hline
             Parameters & Description & Value \\ 
             \hline
             $g_{\omega, -, \text{mb}}^0/2\pi$ (kHz) & moving boundary dispersive coupling & 239 \\ 
             $g_{\omega, -, \text{pe}}^0/2\pi$ (Hz) & photoelastic dispersive coupling & 0.2 \\ 
             ${\lambda}_{-}$ (nm) &  optical mode wavelength & 1494 \\ 
             $\kappa_{-}/2\pi$ (GHz) &  coupling to the environment & 89 \\ 
             \hline
        \end{tabular}
        \caption{FEM modeling and perturbative analysis of moving boundary and photoelastic optomechanical couplings \cite{johnsonPerturbationTheoryMaxwell2002, chanOptimizedOptomechanicalCrystal2012}.}
        \label{tab:fem_mb}
    \end{center}
\end{table*}
\section{Device fabrication}

\subsection{Layer thickness}

\begin{figure*}[h!tbp]
    \centering\includegraphics{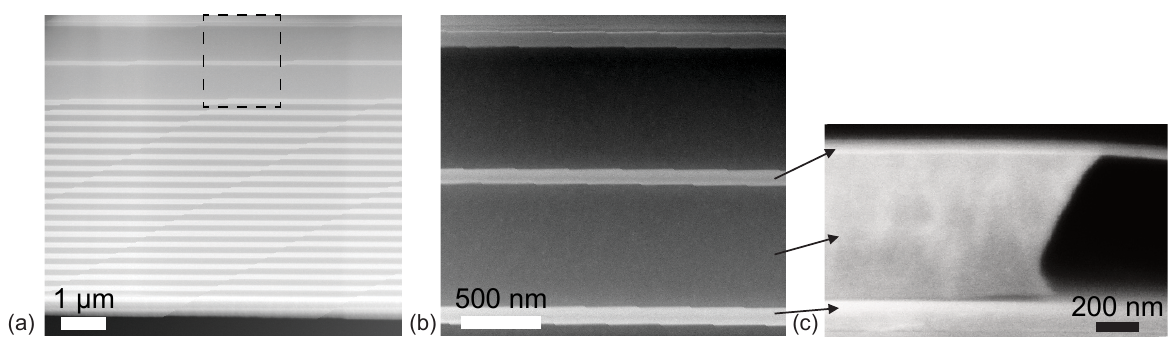}
    \caption{(a) TEM image of the as-grown heterostructure. (b) The inset in (a) is expanded to show the GaAs device and Al$_{0.67}$Ga$_{0.33}$As sacrificial layers. (c) SEM image of a cross-section of the device layer on top of the sacrificial layer after fabrication. The arrows indicate the corresponding layers in (b).
    }
    \label{fig:TEM}
\end{figure*}

Transmission electron microscopy (TEM) and scanning electron microscopy (SEM) were used to determine the thickness of the (Al,Ga)As heterostructure grown by molecular beam epitaxy. The as-grown heterostructure [\fref{fig:TEM}(a)] consists of a DBR mirror with 30 alternating layers of GaAs and Al$_{0.92}$Ga$_{0.08}$As with a targeted reflectivity of 99.999\% at 1480\,nm. The DBR is followed by an Al$_{0.67}$Ga$_{0.33}$As sacrificial layer and a GaAs device layer, which are repeated twice. The devices used in this work were fabricated after stripping the top two layers.

The thickness of the GaAs device layer (Al$_{0.67}$Ga$_{0.33}$As sacrificial layer)  prior to the fabrication process was inferred from the TEM image [\fref{fig:TEM}(a,b)] to be 95\,nm $\pm$ 6.6\,nm (745\,nm  $\pm$ 4.5\,nm). The thickness of the DBR layers was measured to be 102.4 $\pm$ 1.9\,nm for GaAs and  137.6 $\pm$ 2.5\,nm for Al$_{0.92}$Ga$_{0.08}$As.

The SEM image [\fref{fig:TEM}(c)] shows the cross-section of a device after completing all fabrication steps. From this image, the GaAs device layer thickness after fabrication is evaluated to be about 75\,nm$\pm$ 2.4\,nm, which is about 20\,nm thinner than the as-grown layer. Assuming a device layer thickness of 75\,nm in our simulations, we obtain an excellent match between simulated and observed mechanical eigenfrequencies and optical reflectance spectra. The precise reason for the reduction in GaAs layer thickness is presently unknown to us. We can speculate that interdiffusion of Al and Ga in the (Al,Ga)As heterostructure occurred \cite{Beernink1995}. Since the GaAs device layer is sandwiched between two 745\,nm-thick Al$_{0.67}$Ga$_{0.33}$As layers, Al could have diffused into the GaAs device layer. During HF wet etch processing, a small Al content in the GaAs device layer would then have lead to a reduction in layer thickness.

\section{Mechanical characterization}

\subsection{Residual Tensile stress in GaAs device layer}
\label{SM:FreqvsStress}

\begin{figure}[t!bhp]
    \centering
    \includegraphics{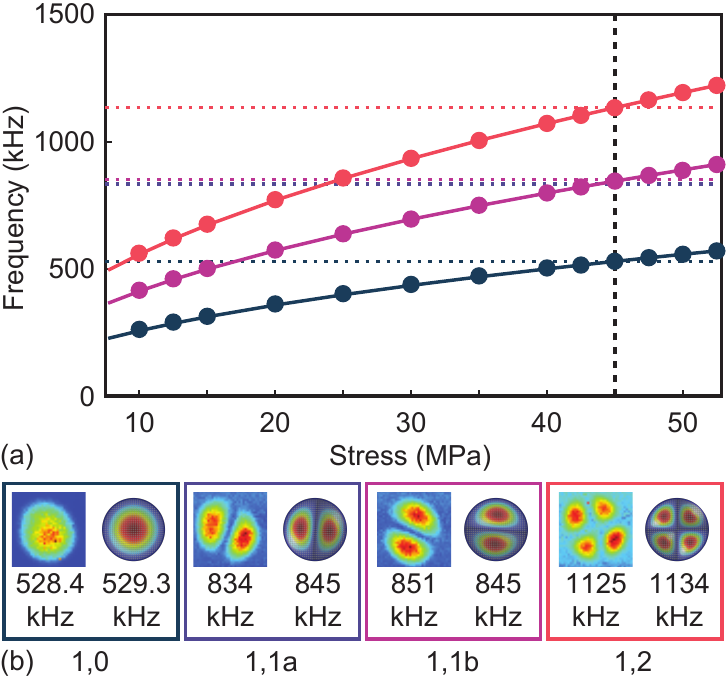}
    \caption{(a) Finite element simulation results for the mechanical frequencies of 100\,µm diameter suspended PhC membrane in a 75\,nm-thick GaAs device layer for varying tensile stress (lines are a guide to the eye). The dotted horizontal lines indicate the measured mechanical frequencies. (b) Mode tomography of the measured mechanical modes seen in (a) along with the simulated mode shape and frequencies. }
    \label{fig:stress}
\end{figure}

We expect some tensile stress in the GaAs device layer due to lattice mismatch between the (Al,Ga)As and GaAs layers as discussed in Ref.~\cite{kini2020suspended}. To estimate the stress, we perform FEM simulations of a 100\,$\mu$m diameter membrane with $r_\PhC$ = 355\,nm and $a_\PhC$ = 1081\,nm in a 75\,nm-thick GaAs device layer with varying tensile stress from 10\,MPa to 52.5\,MPa. We track the evolution of the first four mechanical eigenmodes of the membrane with varying stress, see \fref{fig:stress}. We find a good match between the measured mechanical frequencies and simulated frequencies at a stress of 45\,MPa. This initial stress is assumed in further simulations of variation in mechanical frequency due to change in PhC hole radius seen in the main paper.

\section{Optical characterization}

\subsection{PhC in vacuum}\label{sec:simu PhC}

\begin{figure}[t!hbp]
    \centering\includegraphics{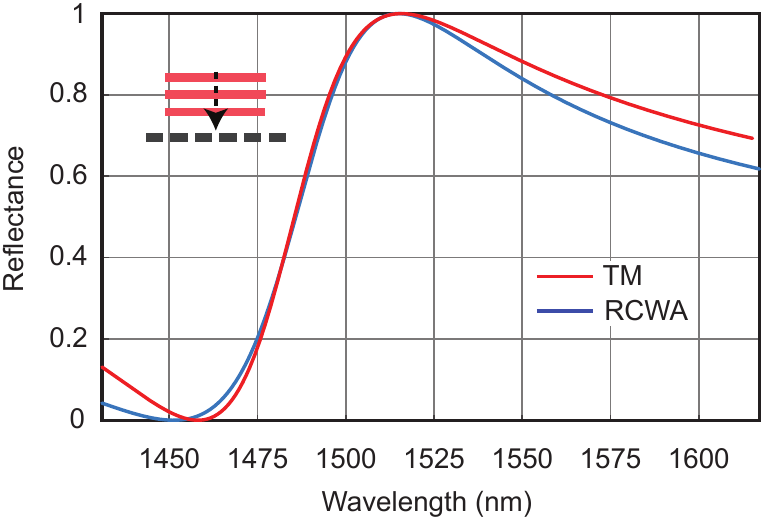}
    \caption{Reflectance of a PhC slab with parameters $d_\PhC = 74$\,nm, $a_\PhC = 1081$\,nm, $r_\PhC = 351$\,nm and $s = 1$. The blue line represents the RCWA simulation and the red line represents the fit obtained from a TM modeling of a single Fano resonance, yielding the parameters $\gamma_d$ and $\lambda_d$.}
    \label{fig:TM_PhC_rg}
\end{figure}

We simulate the reflectance of the PhC slab with PhC holes of radius $r_\PhC$ and lattice constant $a_\PhC$ using RCWA. The result of this RCWA simulation is then fitted to the model of a single Fano resonance as given in \sref{sec:TMModeling} by using \eref{eq:Fano_PhC}. Note that the RCWA simulation captures several Fano resonances that occur in the system, whereby we use in our transfer matrix modelling a single Fano resonance only to simplify the model. The Fano fit yields the parameters $\gamma_d$ and $\lambda_d = 2\pi\,c/\omega_d$ of the Fano resonance. A representative fit is shown in \fref{fig:TM_PhC_rg}.

\subsection{DBR}\label{sec:DBR}

The DBR mirror consists of 30 pairs of alternating layers of 105.9\,nm thick Al$_{0.92}$Ga$_{0.08}$As and 130\,nm GaAs followed by 114.8\,nm quater-wave layer of GaAs and 397\,nm etch-stop layer of Al$_{0.92}$Ga$_{0.08}$As on a GaAs substrate. The GaAs substrate was not back-polished and its back-side roughness results in optical loss. We capture this by modeling the substrate to be an infinite substrate. The reflectance of the DBR mirror was modeled with the transfer matrix method using \eref{mat:int1} and \eref{mat:int2}. 

\begin{figure}[t!bhp]
    \centering\includegraphics{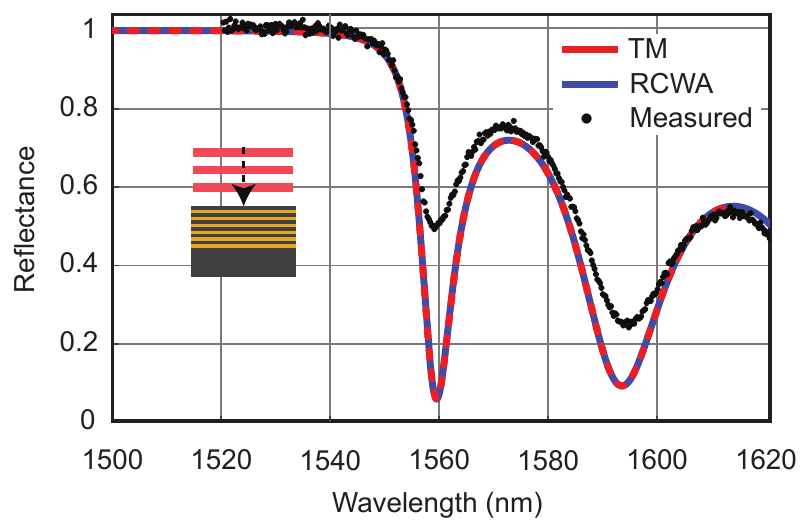}
    \caption{Reflectance of DBR: TM and RCWA modeling compared to experimental data.}
    \label{fig:TM_DBR}
\end{figure}

\begin{figure}[t!bhp]
    \centering\includegraphics{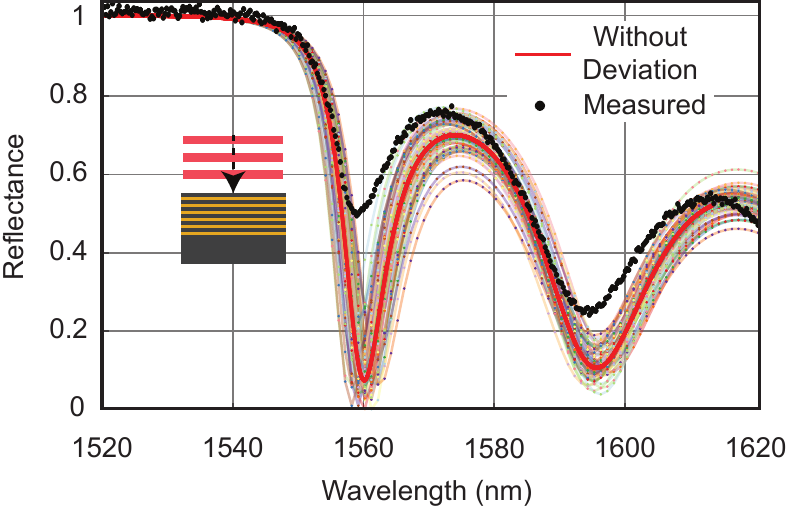}
    \caption{RCWA simulation of the DBR without (red curve) and with thickness variation.}
    \label{fig:RCWA_dbr}
\end{figure}

\fref{fig:TM_DBR} shows the measured reflectance of the DBR along with the simulated reflectance using RCWA and transfer matrix modeling, whereby the latter two yield the same result, as expected. The measurement was performed on the DBR by stripping the top 4 layers of GaAs and (Al,Ga)As. We observe that the experimental data in the minima of the DBR reflectance dips is higher than would be expected from the simulations. 

We attribute this difference to the thickness variation of the individual DBR layers and the finite surface roughness at DBR layer interfaces. From the TEM analysis the estimated thickness variation for the GaAs [(Al,Ga)As] layers of the DBR is 1.5\,nm [0.44\, nm]. This variation along with a finite surface roughness reduces the interference contrast between the electric field reflected of each layer. To illustrate that, we have simulated 50 random sets of DBR mirrors with varying layer thickness given by the above mentioned uncertainties. The result is shown in \fref{fig:RCWA_dbr}. As can be clearly seen, some sets yield an increase in reflectivity at the position of the DBR minima. The remaining discrepancy we attribute to the finite surface roughness at the DBR layer interfaces.

\subsection{Determination of the parameters of the microcavity}\label{sec:microcavparams}

\begin{figure}[h!btp]
    \centering\includegraphics{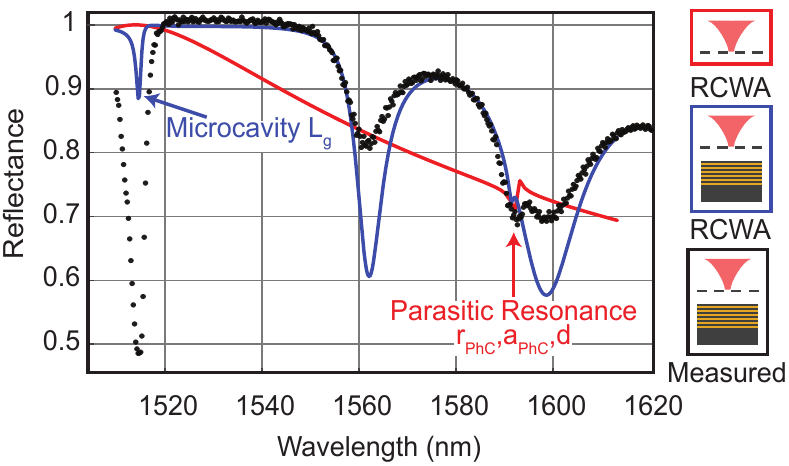}
    \caption{RCWA fitting procedure applied to the experimental data to obtain parameters for modelling the reflectance of the microcavity. We first use the location of the parasitic guided resonance of the PhC to determine the PhC parameters. We then use the location of the microcavity resonance to obtain the cavity length.}
    \label{fig:microcavfitting}
\end{figure}

The microcavity consists of the PhC and the DBR mirror that are separated by a vacuum gap. The reflectance measurement of the microcavity as shown in \fref{fig:microcavfitting} contains features of the individual optical components (PhC and DBR) and of the combined system. We make use of the features of the individual optical components to determine their parameters.

We determine first the properties of the PhC. We extract the radius $r_{\PhC}$ and lattice constant $a_{\PhC}$ of the PhC from SEM analysis. The microcavity spectrum shows a clear signature of the parasitic guided mode resonance of the PhC, in case of  \fref{fig:microcavfitting} it is obtained at a wavelength of 1590\,nm. We then simulate the reflectance of the PhC slab in vacuum using RCWA for varying PhC slab thickness and match the wavelength of the parasitic resonance to obtain the actual thickness $d_{\PhC}$.

We determine the parameters of the DBR mirror independently by measuring the reflectance of the DBR and fitting it to the TM model as described in \sref{sec:DBR}.

Finally, we vary the vacuum gap $L_g$ between the PhC of fixed parameters ($r_{\PhC}$ , $a_{\PhC}$ and $d_{\PhC}$) and the DBR of fixed parameters to match the wavelength of the microcavity resonance, which in \fref{fig:microcavfitting} occurs at 1515\,nm. 

To obtain the parameters for the TM modelling, we first fit the RCWA simulated data for the PhC slab in vacuum to the analytical expression for the reflectivity of a Fano resonance \eref{eq:r_w_zeta} and \eref{eq:Fano_PhC} to obtain $\gd$ and $\omega_d$. The measured microcavity reflectance is fitted to \eref{eq:SM_ReflecTM} to extract $\zeta_2$ and $\Phi$.

\subsection{Simulations of the reflection of the microcavity}

\begin{figure}[t!bhp]
    \centering\includegraphics[width=\linewidth]{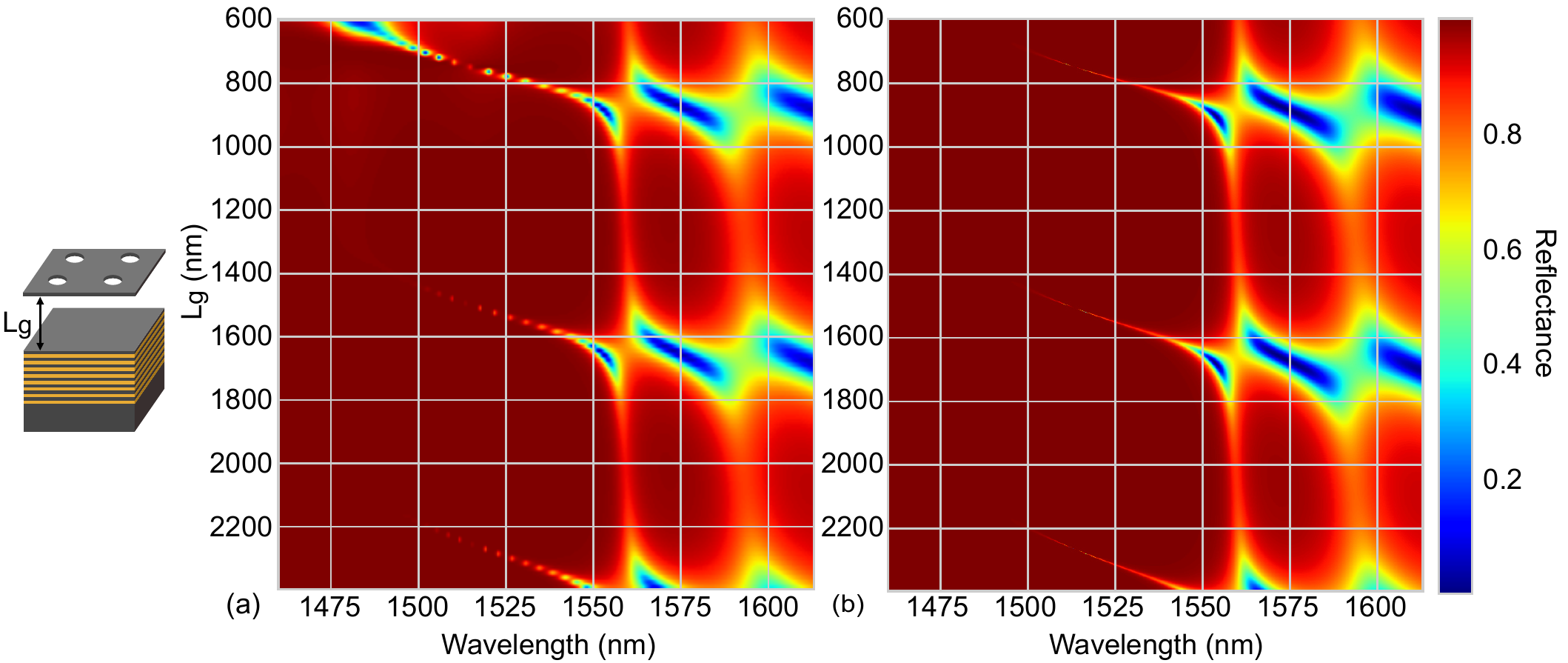}
    \caption{Reflectance map obtained from (a) RCWA simulation and (b) TM modelling.}
    \label{fig:TM_S4_cavity}
\end{figure}

We perform RCWA [\fref{fig:TM_S4_cavity}(a)] and TM simulations [\fref{fig:TM_S4_cavity}(b)] with an effective mirror to obtain the reflectance of the complete system consisting of PhC and DBR that together form the microcavity.

\begin{figure}[t!bhp]
    \centering\includegraphics[width=\linewidth]{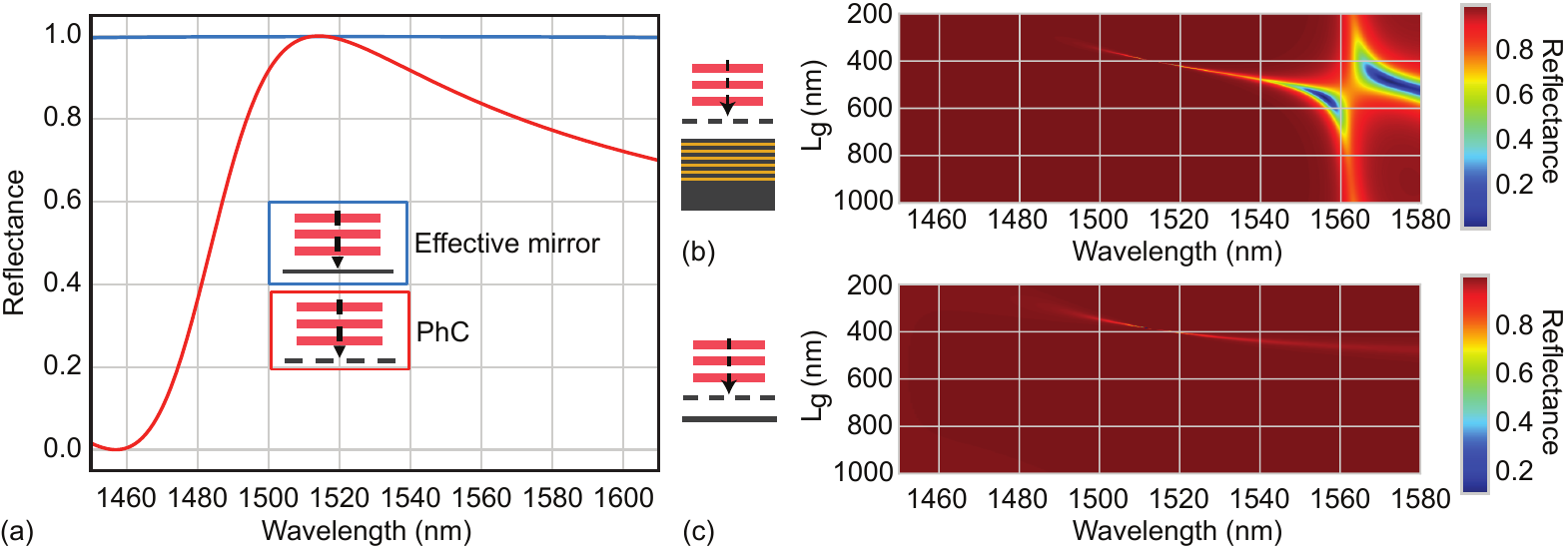}
    \caption{(a) Reflectance with the effective mirror. 
    Reflectance map with the gap variation for (b) with PhC-DBR and (c) PhC and a thin scatterer.}
    \label{fig:TM_mirror}
\end{figure}

When comparing \fref{fig:TM_S4_cavity}(a) and \fref{fig:TM_S4_cavity}(b) with each other, we observe that we can successfully capture the reflectance spectrum at cavity lengths larger than $\lambda/2 = 750\,\text{nm}$. At shorter cavity lengths, near-field effects play a significant role which are not captured in the TM modeling used in our work.

To simplify and represent the DBR reflectivity in terms of polarizability ($\zeta_2$) for TM modelling, we mimic the high reflectivity region of the DBR with the thin scatterer, \fref{fig:TM_mirror}(a). An additional phase was added to the propagation matrix to match the resonance condition of the microcavity with the DBR mirror [\fref{fig:TM_mirror}(b)] to the microcavity with an effective mirror [\fref{fig:TM_mirror}(c)].

\subsection{Beam waist dependent optical measurement of the microcavity}

\begin{figure}[t!bhp]
    \centering\includegraphics{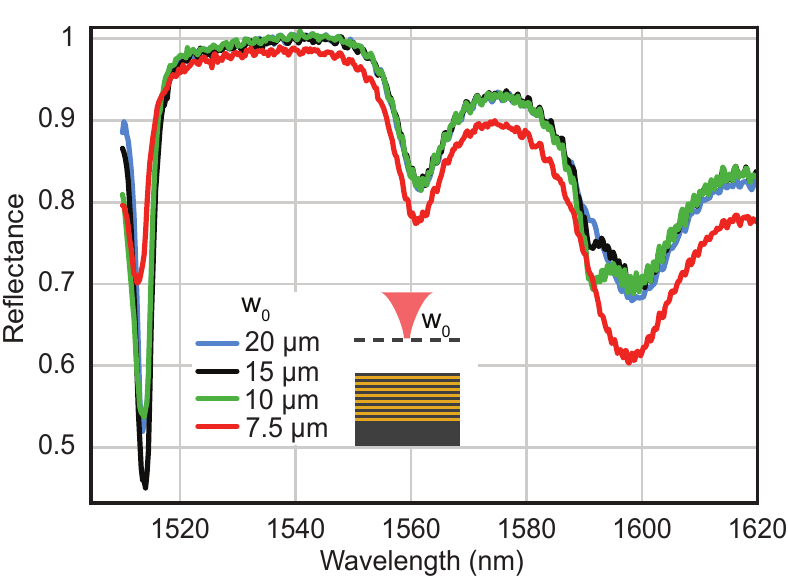}
    \caption{Waist dependence of the microcavity mode and PhC guided mode reflectance.}
    \label{fig:waist}
\end{figure}

In \fref{fig:waist}, we examine the dependence of the microcavity reflectance spectrum on the waist of the incoming beam. We observe that for larger waists, the depth of the microcavity resonance increases. The reason to this effect is that the reflectivity of the PhC slab in vacuum increases for larger waists, see, e.g., Ref.~\cite{kini2020suspended}, which leads to reflectivity mismatch between the PhC and DBR mirror, which determines the microcavity dip depth but does not affect the resonance wavelength. However, we observe that the dip depth is smaller for a waist of 20\,$\mu$m. We attribute this effect to clipping loss due to the finite size of the PhC membrane. We also observe that the parasitic guided resonance of the PhC slab at 1590\,nm becomes narrower and reduces in depth for increasing waist, which had been observed before, e.g., Ref.~\cite{kini2020suspended}.

\section{Optomechanical characterization}

\subsection{Power dependence}

\begin{figure}[b!thp]
    \centering\includegraphics{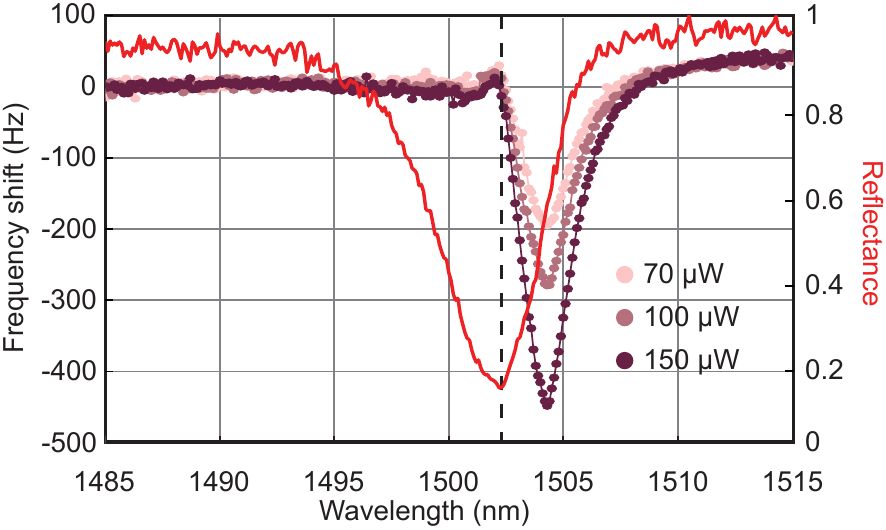}
    \caption{Mechanical resonance frequency shift for different optical powers in dependence of laser wavelength.}
    \label{fig:optomechanics_power}
\end{figure}

\fref{fig:optomechanics_power} shows the power-dependent mechanical frequency shift. The maximal frequency shift increases linearly with optical power.

\begin{figure}[t!bhp]
    \centering\includegraphics[width=\columnwidth]{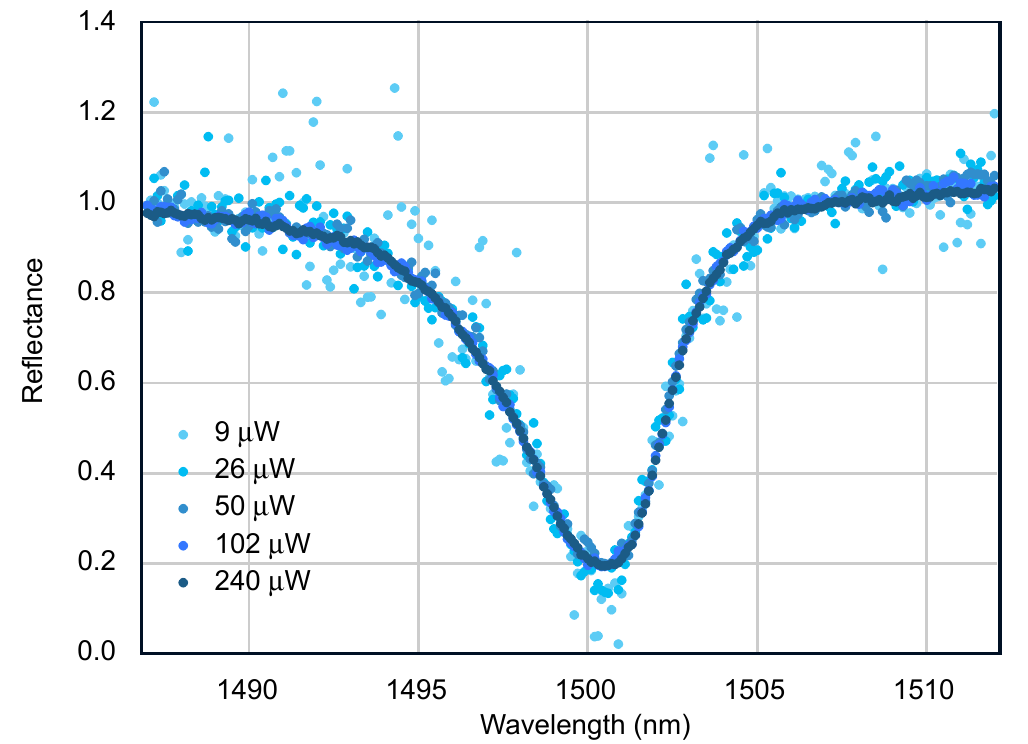}
    \caption{Reflectance of the PhC-DBR microcavity for different optical powers.}
    \label{fig:R_power}
\end{figure}

Moreover, to examine if nonlinear optical effects occur in our microcavity, we measured its reflectance in dependence of optical power, \fref{fig:R_power}. We find that the position and lineshape of the microcavity resonance do not change for the powers used throughout this work. Thus, we exclude nonlinear optical effects to be present in the microcavity when using up to 240\,$\mu$W of input power.

\clearpage
\bibliography{bib_PhCDBR}

\end{document}